# A Market-Inspired Approach for Intersection Management in Urban Road Traffic Networks

**Matteo Vasirani**                                                         matteo.vasirani@urjc.es
**Sascha Ossowski**                                                         sascha.ossowski@urjc.es
*Centre for Intelligent Information Technology*
*University Rey Juan Carlos*
*C/ Tulipán s/n*
*Madrid, 28933, Spain*

## Abstract

Traffic congestion in urban road networks is a costly problem that affects all major cities in developed countries. To tackle this problem, it is possible (i) to act on the supply side, increasing the number of roads or lanes in a network, (ii) to reduce the demand, restricting the access to urban areas at specific hours or to specific vehicles, or (iii) to improve the efficiency of the existing network, by means of a widespread use of so-called Intelligent Transportation Systems (ITS). In line with the recent advances in smart transportation management infrastructures, ITS has turned out to be a promising field of application for artificial intelligence techniques. In particular, multiagent systems seem to be the ideal candidates for the design and implementation of ITS. In fact, drivers can be naturally modelled as autonomous agents that interact with the transportation management infrastructure, thereby generating a large-scale, open, agent-based system. To regulate such a system and maintain a smooth and efficient flow of traffic, decentralised mechanisms for the management of the transportation infrastructure are needed.

In this article we propose a distributed, market-inspired, mechanism for the management of a future urban road network, where intelligent autonomous vehicles, operated by software agents on behalf of their human owners, interact with the infrastructure in order to travel safely and efficiently through the road network. Building on the reservation-based intersection control model proposed by Dresner and Stone, we consider two different scenarios: one with a single intersection and one with a network of intersections. In the former, we analyse the performance of a novel policy based on combinatorial auctions for the allocation of reservations. In the latter, we analyse the impact that a traffic assignment strategy inspired by competitive markets has on the drivers' route choices. Finally we propose an adaptive management mechanism that integrates the auction-based traffic control policy with the competitive traffic assignment strategy.

## 1. Introduction

Removing the human driver from the control loop through the use of autonomous vehicles integrated with an intelligent road infrastructure can be considered as the ultimate, long-term goal of the set of systems and technologies grouped under the name of Intelligent Transportation Systems (ITS). Autonomous vehicles are already a reality. For instance, three *DARPA Grand Challenges*[1] have been held so far. The teams participating in the latest event, the *DARPA Urban Challenge*, competed to build the best autonomous vehi-cles, capable of driving in traffic, performing complex manoeuvres such as merging, passing, parking and negotiating with intersections. The results have shown for the first time that autonomous vehicles can successfully interact with both manned and unmanned vehicular traffic in an urban environment. Several car-makers expect the technology to be affordable (and less obtrusive) in about a decade[2]. Another initiative that fosters this vision is *Connected Vehicle*[3], which promotes research and development of technologies that link road vehicles directly to their physical surroundings, i.e., by vehicle-to-infrastructure wireless communications. The advantages of such an integration span from improved road safety to a more efficient operational use of the transportation network. For instance, vehicles can exchange critical safety information with the infrastructure, so as to recognise high-risk situations in advance and therefore to alert drivers. Furthermore, traffic signal systems can communicate signal phase and timing information to vehicles to enhance the use of the transportation network.

---

1. http://archive.darpa.mil/grandchallenge/

In this regard, some authors have recently paid attention to the potential of a tighter integration of autonomous vehicles with the road infrastructure for future urban traffic management. In the reservation-based control system (Dresner & Stone, 2008), an intersection is regulated by a software agent, called *intersection manager* agent, which assigns reservations of space and time to each autonomous vehicle intending to cross the intersection. Each vehicle is operated by another software agent, called *driver* agent. When a vehicle is approaching an intersection, the driver requests that the intersection manager reserve the necessary space-time slots to safely cross the intersection. The intersection manager, provided with data such as vehicle ID, vehicle size, arrival time, arrival speed, type of turn, etc., simulates the vehicle's trajectory inside the intersection and informs the driver whether its request is in conflict with the already confirmed reservations. If such a conflict does not exist, the driver stores the reservation details and tries to meet them; otherwise it may try again at a later time. The authors show through simulations that in situations of balanced traffic, if all vehicles are autonomous, their delays at the intersection are drastically reduced compared to traditional traffic lights.

In this article we explore how different lines of research in artificial intelligence and agent technology can further improve the effectiveness and applicability of Dresner and Stone's approach, assuming that all vehicles are autonomous and capable of interacting with the regulating traffic infrastructure. We extend the reservation-based model for intersection control at two different levels.

### 1.1 Single Intersection

For a single intersection, our objective is to elaborate a new policy for the allocation of reservations to vehicles that takes into account the drivers' different attitudes regarding their travel times. Instead of granting the disputed resources (intersection space and time) to the first agent that requests them, we intend to allocate them to the agents that value them the most, while maintaining an adequate level of efficiency and fairness of the system. Our main contribution in this regard is the definition of an auction-based allocation policy for assigning reservations. This policy models incoming requests as bids over an intersection's available space-time slots and tries to maximise the overall value of the accepted bids. Due to the combinatorial nature of the auction and the restrictions of our scenario (mainly real-time execution and safety), we define a specific auction protocol, adapt an algorithm for winner determination for our purposes, and evaluate the behaviour of the approach.

---

2. See for example Alan Taub, General Motors Vice President of Global R&D, at the 18th World Congress on Intelligent Transport Systems, October 17th, 2011.
3. http://www.its.dot.gov/connected_vehicle/connected_vehicle.htm

### 1.2 Network of Intersections.

To extend Dresner and Stone's approach to a network of intersections, we focus on the problem of *traffic assignment*, conceived as a distributed choice problem where intersection managers try to affect the decision making of the driver agents. In particular, we use *markets* as mediators for our distributed choice and allocation problem (Gerding, McBurney, & Yao, 2010). Our contribution to the attainment of the above objective is twofold. First, we build a computational market where drivers must acquire the right to pass through the intersections of the urban road network, implementing the intersection managers as competitive suppliers of reservations which selfishly adapt the prices to match the actual demand. Second, we combine the competitive strategy for traffic assignment with the auction-based control policy at the intersection level into an adaptive, market-inspired, mechanism for traffic management of reservation-based intersections.

The article is structured as follows. Section 2 provides an overview of the use of artificial intelligence and agent technology in the field of ITS. In Section 3 we briefly review the key elements of the reservation-based intersection control model that our work sets out from. In Section 4 we present our policy for the allocation of reservations at a single intersection, inspired by combinatorial auction theory. In Section 5 we extend the reservation-based model to network of intersections. Finally, we conclude in Section 6.

## 2. Related Work

To achieve the goals pursued by the ITS vision there is an increasing need to understand, model, and govern such systems at both the individual (micro) and the societal (macro) level. Transportation systems may contain thousands of autonomous entities that need to be governed, which raises significant technical problems concerning both efficiency and scalability. The inherent distribution of traffic management and control problems, their high degree of complexity, and the fact that the actors in traffic and transportation systems (driver, pedestrians, infrastructure managers, etc.) fit the concept of autonomous agent very well, allow for modelling ITS in terms of agents that interact so as to achieve their goals, selfishly as well as cooperatively. Therefore, traffic and transportation scenarios are extraordinarily appealing for multiagent technology (Bazzan & Klügl, 2008). In this section, we outline some key dimensions of ITS and briefly review relevant literature on the use of artificial intelligence and multiagent techniques in the field.

### 2.1 Traffic Control and Traffic Assignment

Traffic control refers to the regulation of the access to a disputed road transport resource. Traffic control systems manage traffic along arterial roadways, employing traffic detectors, traffic signals, and various means of communicating information to drivers. Freeway control systems manage traffic along highways, employing traffic surveillance systems, traffic control measures on freeway entrance ramps (ramp metering), and lane management.

Traffic control at intersections, based on traffic lights, is the major control measure in urban road networks. This type of control typically applies off-line optimisation on the basis of historical data. TRANSYT (Robertson, 1969) is a well-known and frequently applied signal control strategy, but it cannot adapt dynamically to changing demand patterns. Other control techniques, such as SCOOT (Hunt, Robertson, Bretherton, & Winton, 1981), use real-time traffic volume rather than historical data to run optimisation algorithms and compute the optimal signal plan.

Traffic assignment refers to the problem of the distribution of traffic in a network, considering demands between several locations, and the capacity of the network. In general, demand may change in a non-predictable way, due to changing environmental conditions, exceptional events, or accidents. This, in turn, leads to under-utilisation of the overall network capacity, whereby some links are heavily congested while capacity reserves are available on alternative routes. To address this problem, different traffic management techniques, involving information broadcast as well as control and optimisation, can be employed. For example, route guidance and driver information systems (RGDIS) may be employed to improve the network efficiency via direct or indirect recommendation of alternative routes (Papageorgiou, Diakaki, Dinopoulou, Kotsialos, & Wang, 2003). These communication devices may be consulted by a potential road user to make a rational decision regarding whether or not to carry out (or postpone) the intended trip, the choice of transport mode (car, bus, underground, etc.), the departure time selection and the route choice.

Traffic control and assignment have different focuses and can therefore be combined into a single management policy that takes explicitly into account the mutual interactions between signal control policies and user route choices (Meneguzzer, 1997).

### 2.2 Isolated and Coordinated Traffic Control

Most traffic control strategies use control devices (e.g., traffic lights, variable message signs, ramp meters) and surveillance devices (e.g., loop detectors, cameras) to manage a physical traffic network. In isolated control, only a small portion of the network (e.g. a single intersection) is modelled, and techniques from control theory are employed to determine signal cycles so as to minimise the vehicles' total delay. For instance, da Silva *et al.* proposed a reinforcement learning system for traffic lights that copes with the dynamism of the environment by incrementally building new models of the environmental state transitions and rewards (da Silva, Basso, Bazzan, & Engel, 2006). When the traffic pattern changes, an additional model is created and a new traffic signal plan is learned. The creation of new models is controlled by a continuous evaluation of the prediction errors generated by each partial model.

In coordinated control, the settings of several control devices are adapted to each other, so as to achieve a smooth traffic flow at the network level (i.e., "green waves") rather than at a single intersection. By allowing the individual devices to coordinate their actions based on the information they receive from sensors and from each other, coherent traffic control plans are often generated faster and more accurately compared to a human traffic operator (van Katwijk, Schutter, & Hellendoorn, 2009). For instance, distributed constraint optimisation (DCOP) techniques have recently been applied to the coordination of control devices (Junges & Bazzan, 2008). Each traffic signal agent is assigned to one or several variables of the DCOP, which have inter-dependencies and conflicts (e.g., two neighbouring intersections giving preference to different directions of traffic.). A mediator agent is in charge of resolving these conflicts when they occur, recommending values for the variables associated to the agents involved in the mediation.

### 2.3 Time Perspective

The time perspective refers to the stage in which the decision-making process of an ITS application takes place. *Operational* decision-making in ITS refers to short term issues, such as controlling traffic at an intersection. *Tactical* decision-making deals with medium-term issues, such as anticipating congestion by diverting traffic on different routes or influencing demand patterns. Finally, *strategic* decision-making typically involves long-term decisions, e.g. planning the construction of new roads, highways or parking hubs.

Many AI-based ITS partially automate the operational part of road traffic control tasks. Tactical and strategic decision-making is still mainly a human activity (e.g., carried on by city planners). Some more recent decision-support systems address tactical questions as well. $InTRYS$ (Hernández, Ossowski, & García-Serrano, 2002), for instance, is a multiagent system aimed at assisting operators in a traffic control centre to manage an urban motorway network. The system is capable of engaging in dialogues with the operators, e.g. to diagnose the causes of detected traffic problems, to construct coherent sets of driver information messages, and to simulate the expected effects of such control plans.

### 2.4 Information to Drivers

Cooperative systems can improve dynamic routing and traffic management (Adler, Satapathy, Manikonda, Bowles, & Blue, 2005), using information services aimed at giving advice to drivers and efficiently assigning traffic among the network. This is a difficult problem as collective route choice performed by selfish agents often leads to equilibrium strategies that are far from social welfare optima (Roughgarden, 2003). Providing information about the congestion of links or sharing partial views of vehicle choices, as in context-aware routing (Zutt, van Gemund, de Weerdt, & Witteveen, 2010), may improve the system's efficiency.

### 2.5 Domain Knowledge

Domain and topological knowledge can be exploited to structure both the architecture and the reasoning models of ITS. For instance, Choy *et al.* propose a cooperative, hierarchical, multiagent system for real-time traffic signal control (Choy, Srinivasan, & Cheu, 2003). The control problem is divided into various sub-problems, each of them handled by an intelligent agent that applies fuzzy neural decision-making. The multiagent system is hierarchical, since decisions made by lower-level agents are mediated by their respective higher-level agents. The $InTRYS$ system (Hernández et al., 2002) conceives the traffic dynamics in terms of so-called *problem areas*, which are defined based on the expertise of traffic engineers. Each problem area is controlled by a separate traffic control (software) agent. Knowledge modelling and reasoning techniques are applied to integrate local control strategies (proposed by the different traffic control agents) into a coherent global plan for the whole traffic network.

### 2.6 Learning and Adaptation

ITS often rely on learning techniques to adapt to changing or unknown traffic conditions. For instance, traffic light agents may use reinforcement learning to minimise the overall waiting time of vehicles (Steingrover, Schouten, Peelen, Nijhuis, & Bakker, 2005; Wiering, 2000). The control objective is global, although actions are local to the agents. The state of the learning task is represented as an aggregation of the waiting times of individual vehicles at the intersection. Traffic light agents learn a value function that estimates expected waiting times of vehicles given different settings of traffic lights.

Several authors focus on self-organising and self-adapting mechanisms for traffic control (Gershenson, 2005; Lämmer & Helbing, 2008), where traffic lights self-organise with no direct communication between them. The local interactions between neighbouring traffic lights lead to emergent coordination patterns such as "green waves". In this way, an efficient, decentralised traffic light control is achieved, as a combination of two rules, one that aims at optimising the flow and one that aims at stabilising it. In the $TRYSA_2$ system (Hernández et al., 2002), traffic agents use a mechanism called *structural cooperation* (Ossowski & García-Serrano, 1999) to locally coordinate their signal plan proposals without the need to rely on dedicated domain (coordination) knowledge.

### 2.7 Market-Based Coordination

Being a complex system, traffic is well suited for the application of market-based coordination mechanisms at different levels. These mechanisms replicate the functioning of real markets (i.e., auctions, bargaining, etc.) in order to coordinate the activities and goals pursued by a set of agents. The agents that regulate the infrastructure can be built to act as a team, i.e., they may share a global objective function that represents the system designer's preferences over all possible solutions, as it occurs in multi-robot domains (Dias, Zlot, Kalra, & Stentz, 2006). In line with this perspective, Vasirani and Ossowski (2009b) proposed a market-based policy for traffic assignment. The authors put forward a cooperative learning model so as to coordinate the prices of several intersections. The experimental results showed that, in general, an increase in the profit raised by a team of intersections is aligned with reduced average travel times. A limitation of this work is the number of interactions with the environment that are required in order for the price vector that maximises overall profit to be learned.

If we extend the focus to include selfish driver agents and their interaction with the infrastructure agents, a non-cooperative scenario arises. For instance, an auction-based policy for intersection control is proposed in the work of Schepperle and Bohm (2007). In this work, an intersection controlled by an intelligent agent starts an auction for the earliest time slot among the vehicles that are approaching the intersection on each lane. The authors assume that the agent that controls an intersection can detect if an approaching vehicle has another vehicle in front of it. In this case, the former is not allowed to participate in the auction (i.e., its bids are not processed), so as to ensure that only vehicles that do not have physical impediments to cross the intersection are allowed to participate in the auction.

Furthermore, since a non-combinatorial auction is run to allocate the earliest time-slot, only one bidder (i.e., driver) is entitled to get a specific time-slot, which can lead to inefficiencies in the assignments.

The field of transport economics also studies the allocation of resources used to move road users from place to place (Small & Verhoef, 2007). However, it follows a more static and analytical approach that requires extensive knowledge of supply and demand functions. Such information is often hard to obtain and extract, so usually findings from the field are hard to transfer directly to ITS.

## 2.8 Discussion

In this work, we mainly focus on the *operational* time perspective, since our aim is to manage an advanced traffic infrastructure that regulates the route choices of autonomous vehicles, while tactical and strategic decisions are left to the human users. In order to make the proposed mechanisms broadly adoptable, we minimise the *domain knowledge* necessary to set up our models. While the software agents that reside in the traffic management infrastructure need to be aware of the remaining infrastructure agents, they do not require expert knowledge related to the underlying traffic system. We focus on *local adaptation* mechanisms, rather than learning techniques, to enforce emergent coordination among the software agents that reside in the traffic management infrastructure. Furthermore, we put forward a *market-based coordination* framework that involves both the infrastructure and the drivers. The infrastructure agents coordinate their actions in an indirect way as competitive market participants that aim to match supply with demand. The driver agents participate in the allocation of the road network capacity through an auction-based mechanism that regulates the assignment of the right to cross an intersection. Finally, we recognise the importance of providing *information to drivers* in order to influence their decision making. In particular, we assume the existence of propagation mechanisms, so that the market price information is available to the drivers, thus potentially influencing their collective behaviour[4].

## 3. Reservation-Based Intersection Control

The applications of AI techniques and multiagent technology in the traffic domains that were detailed in the previous section conceive that the ITS lies in the infrastructure and its components (traffic lights, message signs, sensors, etc.), while the vehicles are usually treated as particles of a traffic flow that a control policy cannot individually address. Nevertheless, the continuous advances in software and hardware technologies will make a tighter integration between vehicles and infrastructure possible. Even today, vehicles can be equipped with features such as cruise control (Ioannou & Chien, 1993) and autonomous steering (Krogh & Thorpe, 1986). Small-scale systems of autonomous guided vehicles (AGV) already exist, for example in factory transport systems. If this trend continues, one day fully autonomous vehicles will populate our road networks. In this case, given that the system will comprise a variable (and possibly huge) number of vehicles, an open infrastructure is needed to control and schedule the transit of AGVs. In fact, nowadays centralised AGV control systems know the number of the vehicles, their origins and destinations, before the route planning takes place. In the case of an urban road traffic scenario, such an approach is certainly unfeasible.

---

4. Setting up such "price index boards" is technically feasible already today: for instance, the NYSE indexes approximately 8500 stocks, whose price variations are spread worldwide almost immediately.

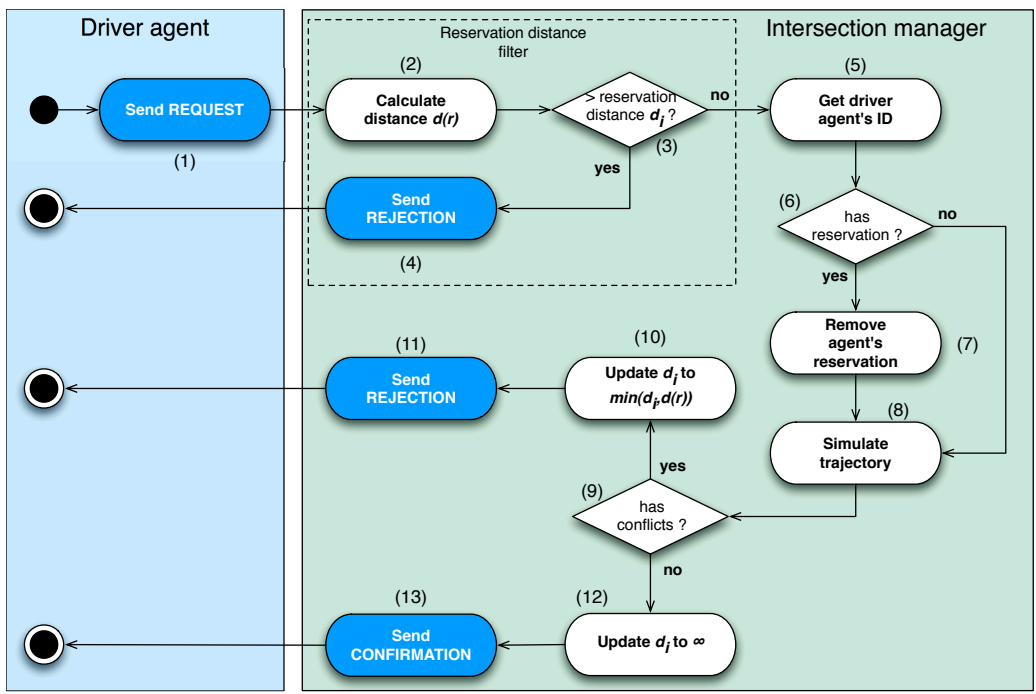

Figure 1: Reservation-based protocol with FCFS policy

In this section we present some details of the reservation-based system for intersection control (Dresner & Stone, 2008) that are relevant for this work[5]. In particular we outline the policy executed by intersection managers to process reservations requests (Section 3.1) and analyse the impact that the distance at which the reservation is sent has on the performance of the control mechanism (Section 3.2).

## 3.1 Protocol

The reservation-based control system proposed by Dresner and Stone assumes the existence of two different kinds of software agents: intersection manager agents and driver agents. The intersection manager agent controls the space of an intersection and schedules the crossing of each vehicle. The driver agent is the entity that autonomously operates the vehicle (in the following we will use the terms "intersection manager" and "driver" for short, to refer to the software agents that control an intersection and a vehicle respectively). The protocol, using the *first-come-first-served* policy (FCFS), is summarised in Figure 1. Each driver,

---

5. We remark that in this work we engineered the *basic* aspects of the reservation-based system. We did not consider more advanced features, such as acceleration within the intersections, safety buffers or edge tiles. The basic functioning of the reservation-based intersection that we assume in this work is the same in every experimental scenario that we compare. In this way a fair comparison between different policies for the allocation of reservations is guaranteed.

```
(request reservation
        :sender    D-3548
      :receiver    IM-05629
      :content(
                 :arrival_time    08:03:15
                 :arrival_speed   23km/h
                 :lane            2
                 :type_of_turn    LEFT
              )
)
```

Figure 2: Example of a REQUEST message

when approaching the intersection, contacts the intersection manager by sending a REQUEST message (1). The message contains the vehicle's ID, the arrival time, the arrival speed, the lane occupied by the vehicle in the road segment that approaches the intersection and the intended type of turn (see Figure 2 for an example of REQUEST message). The intersection manager calculates the distance $d(r)$ from which the driver is sending the reservation request $r$ (2). If the distance is greater than the maximum reservation distance $d_i$ of the lane that the driver is occupying (3), the request is rejected without processing it (4). Otherwise, the intersection manager proceeds to evaluate whether it can be accommodated or not. First, the driver's ID is parsed (5), and if the driver already has a prior reservation (6), this reservation is removed (7). Then, with the information contained in the REQUEST message, the intersection manager simulates the vehicle trajectory, calculating the space needed by the vehicle over time in order to check if there are potential conflicts (8). If so (9), the intersection manager updates the maximum reservation distance $d_i$ (10) and replies with a REJECTION message (11). Otherwise, the maximum reservation distance $d_i$ is updated to infinite (12) and the intersection manager replies with a CONFIRMATION message (13), which implies that the driver's request is accepted.

The FCFS policy implies that if two drivers send requests that require the same space-time slots inside the intersection, the driver that sends the request first will obtain the reservation. In extreme cases this policy is clearly inefficient. Consider the case of a set of $n$ vehicles, $v_1, v_2, \ldots, v_n$, such that $v_1$'s request has conflicts with every other vehicle, but that $v_2, \ldots, v_n$ do not have conflicts with one another. If $v_1$ sends its request first, it will be granted and all other vehicles' requests will be rejected. On the other hand, if it sends its request last, the other $n-1$ vehicles will have their requests confirmed, whilst only $v_1$ will have to wait. Nevertheless, FCFS has the advantage of being a simple policy, which only needs the minimum amount of information necessary to implement a reservation-based intersection control.

### 3.2 Reservation Distance

The protocol detailed above would be prone to deadlock situations, if it did not make use of the reservation distance filter. Consider two vehicles, A and B, with A moving in front of B (see Figure 3). Suppose also that B cannot safely overtake A. If A and B send a request for the same space-time slots inside the intersection, the first request that the intersection manager receives will be accepted, and the second one will be rejected. If vehicle B, which is behind vehicle A, obtains the reservation, the result will be that vehicle A is not able to cross because it does not hold a confirmed reservation. This in turn prevents vehicle B making use of its reservation. If vehicle B always sends its request first, then a deadlock situation arises, with vehicle A physically blocking vehicle B, and vehicle B blocking vehicle A by getting the disputed reservation.

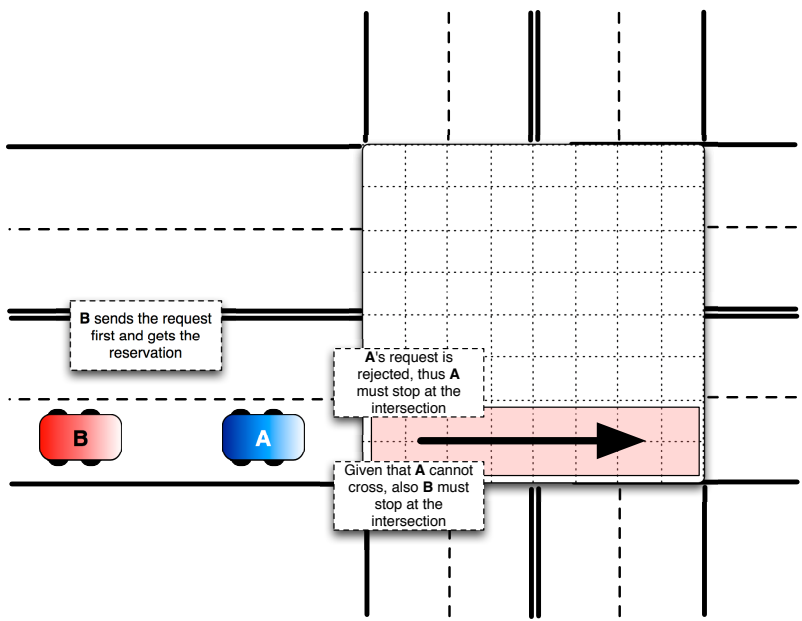

Figure 3: Potential deadlock situation.

To avoid the occurrence of these deadlock situations, Dresner and Stone proposed the use of the reservation distance as a heuristic criterion for filtering out reservation requests that could generate deadlock situations. Since the drivers communicate the time at which they plan to arrive at the intersection, as well as what their speed will be when they get there (quantities which the drivers have no incentive to misrepresent), it is possible to approximate a vehicle's distance from the intersection, given a reservation request by that vehicle. This heuristic approximation, called the reservation distance $d(r)$, is calculated as $d(r) = v_a \cdot (t_a - t)$, where $v_a$ is the proposed arrival speed of the vehicle, $t_a$ is the proposed arrival time of the vehicle, and $t$ is the current time.

This approximation assumes that the vehicle is maintaining a constant speed. The reservation processing policy uses it as follows. For each lane $i$, the policy has a variable $d_i$, initialised to infinity, that represents the maximum distance from which a driver can send a reservation request. For each reservation request $r$ from lane $i$, the policy computes the reservation distance, $d(r)$. If $d(r) > d_i$, $r$ is rejected. If, on the other hand, $d(r) \leq d_i$, $r$ is processed as normal. If $r$ is rejected after being processed as normal, $d_i \leftarrow min(d_i, d(r))$. Otherwise, $d_i \leftarrow \infty$. While the use of the reservation distance does not guarantee that mutually blocking situations never occur, it does prevent these situations from degenerating into deadlocks.

## 4. Single Intersection

For a single reservation-based intersection, the problem that the intersection manager has to solve is allocating the reservations among a set of drivers in a way that a specific objective is maximised. This objective can be, for instance, minimising the *average* delay caused by the presence of the regulated intersection. In this case, the simplest policy to adopt is allocating a reservation to the first agent that requests it, as occurs with the FCFS policy proposed by Dresner and Stone in their original work. Another work in line with this objective takes inspiration from adversarial queuing theory for the definition of several alternative control policies that aim at minimising the average delay (Vasirani & Ossowski, 2009a)

However, these policies ignore the fact that in the real world, depending on the context and their personal situation, people value the importance of travel times and delays quite differently. Since processing the incoming requests to grant the associated reservations can be considered as the process of assigning resources to agents that request them, one may be interested in an intersection manager that aims to allocate the disputed resources to the agents that value them the most. In line with approaches from mechanism design, we assume that the more a human driver is willing to pay for the desired set of space-time slots, the more they value the good. Therefore, we rely on combinatorial auction theory (Krishna, 2002) for the definition of an auction-based policy for the allocation of resources.

### 4.1 Auction-Based Policy

To formalise an auction-based policy for processing incoming reservation requests, it is necessary to specify the auction design space. This includes the definition of the disputed resources, the rules that regulate the bidding and the clearing policy.

#### 4.1.1 AUCTIONED RESOURCES

The first step for the design of any auction is the definition of the resources (or items) to be allocated. The nature of items determines which type of auction can be employed to allocate them. In our scenario, the auctioned good is the use of the space inside the intersection at a given time. We model an intersection as a discrete matrix of space slots. Let $\mathcal{S}$ be the set of the intersection space slots, $\mathcal{S} = \{s_1, s_2, \ldots, s_m\}$. Let $t_{now}$ be the current time, and $\mathcal{T} = \{t_{now} + \tau, \forall \tau \in \mathbb{N}\}$ the set of future time-steps. The set of items that a bidder can bid for is the set $\mathcal{I} = \mathcal{S} \times \mathcal{T}$. Due to the nature of the problem, a bidder is only interested in bundles of items over the set $\mathcal{I}$. In the absence of acceleration in the intersection, a reservation request (Figure 4) implicitly defines which space slots at which time the driver needs in order to pass through the intersection[6]. Thus, the items must necessarily be allocated through a combinatorial auction.

---

6. This computation is easily done by the intersection manager, which knows the geometry of the intersection. If the vehicles were to calculate the trajectory, they would need to know the geometry of every intersection they pass through.

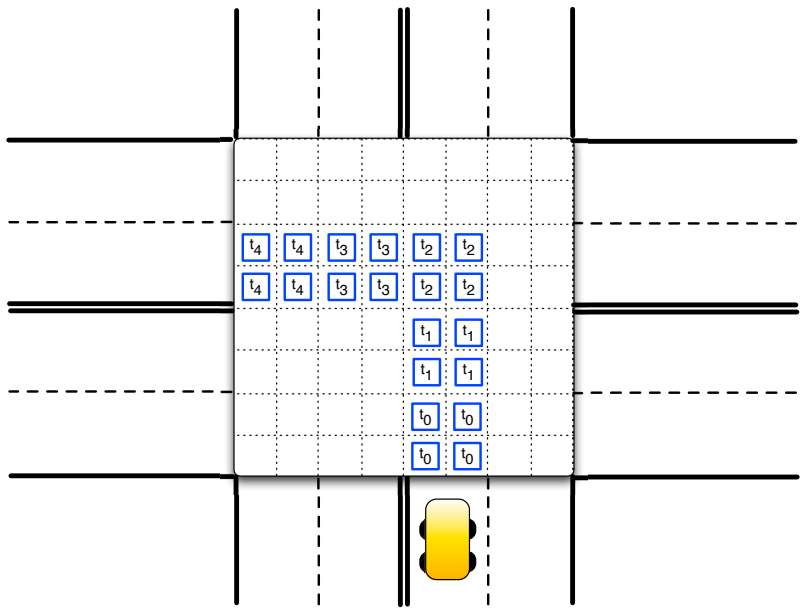

Figure 4: Bundle of items defined by a reservation request.

4.1.2 Bidding Rules

The bidding rules define the form of a valid bid accepted by the auctioneer (Wurman, Wellman, & Walsh, 2001). In our scenario, a bid over a bundle of items is implicitly defined by the reservation request. Given the parameters *arrival time*, *arrival speed*, *lane* and *type of turn*, the auctioneer (i.e., the intersection manager) is able to determine which space slots are needed at which time. Thus, the additional parameter that a driver must include in its reservation request is the *value of its bid*, i.e., the amount of money that it is willing to pay for the requested reservation.

A bidder is allowed to withdraw its bid and to submit a new one. This may happen, for instance, when a driver that submitted a bid $b$, estimating to be at the intersection at time $t$, realises that, due to changing traffic conditions, it will more likely to be at the intersection at time $t + \Delta t$, thus making the submitted bid $b$ useless for the driver. In this case the driver has no guarantees of safety regarding its crossing of the intersection. Thus, the rational thing to do in this case, as the driver would not want to risk being involved in a car accident, is resubmitting the bid with the updated arrival time. However, the new bid must be greater than or equal to the value of the previous one. This constraint avoids the situation whereby a bidder "blocks" one or several slots for itself, by acquiring them early and with overpriced bids. Even though this would oblige others to try to reserve alternative slots, and thus make the desired slot less disputed, the bidder cannot take advantage of this, as it cannot withdraw its initial bid and resubmit lower bids in order to obtain the same reservation at a lower price.

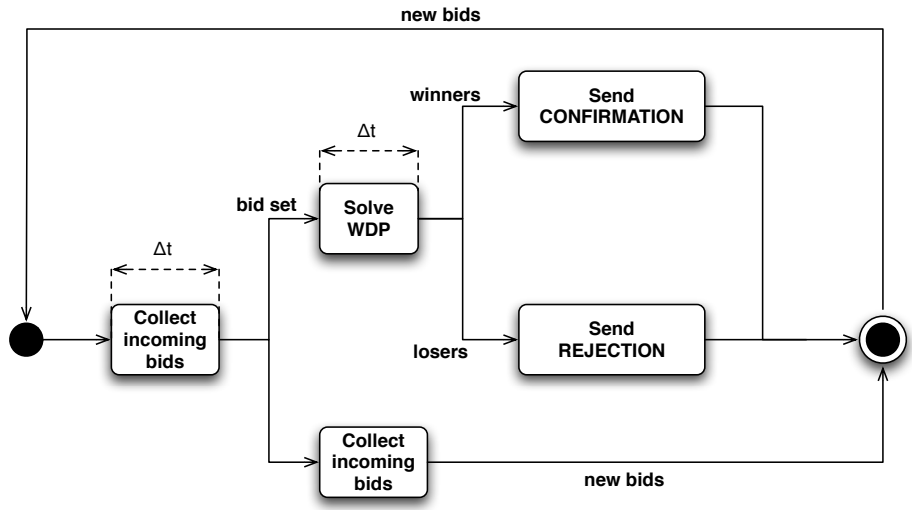

Figure 5: Auction policy

### 4.1.3 AUCTION POLICY

The auction policy (see Figure 5) starts with the auctioneer waiting for bids for a certain amount of time $\Delta t$. Once the new bids are collected, they constitute the bid set. Then, the auctioneer executes the algorithm for the winner determination problem (WDP), and the winner set is built, containing the bids whose reservation requests have been accepted. During the WDP algorithm execution, the auctioneer still accepts incoming bids, but they will only be included in the bid set of the next round. Then the auctioneer sends a CONFIRMATION message to all bidders that submitted the bids contained in the winner set, while a REJECTION message is sent to the bidders that submitted the remaining bids. Then a new round begins, and the auctioneer collects new incoming bids for a certain amount of time[7].

### 4.1.4 WINNER DETERMINATION ALGORITHM

Since the auction must be performed in real-time, both the bid collection and the winner determination phase must be time-bounded, that is, they must occur within a specific time window. This implies that optimal and complete algorithms for the WDP (Leyton-Brown, Shoham, & Tennenholtz, 2000; Sandholm, 2002) are not suited for this kind of auction. An algorithm with *anytime* properties is needed (Hoos & Boutilier, 2000), so that the longer the algorithm keeps executing, the better the solution it finds.

---

7. For safety reasons the auctioneer cannot spend too much time collecting bids, nor can it deallocate previously granted reservations. Therefore it is possible that a low-valued bid, in the winner set at round $k$, impedes the allocation of the disputed reservation to some high-valued bids, submitted at round $k+n$. In this case, the second bidder should slow down and resubmit a new (possibly winning) bid. Although in theory the bid-delay relation (Figure 7) could be worsened by the unrelated sequence of auctions, in practice the effect is negligible.

**Algorithm 1** Winner determination algorithm
---
$\mathcal{B} \leftarrow allBids$
$\mathcal{W} \leftarrow \emptyset$
$start \leftarrow currentTime$
**while** $currentTime - start < 1\ sec$ **do**
  $\mathcal{A} \leftarrow \emptyset$
  **for** $step = 1$ to $|\mathcal{B}|$ **do**
    $step \leftarrow step + 1$
    $random \leftarrow drawUniformDistribution(0, 1)$
    **if** $random < wp$ **then**
      $b \leftarrow selectRandomlyFrom(\mathcal{B} \setminus \mathcal{A})$
    **else**
      $highest \leftarrow selectHighestFrom(\mathcal{B} \setminus \mathcal{A})$
      $secondHighest \leftarrow selectSecondHighestFrom(\mathcal{B} \setminus \mathcal{A})$
      **if** $highest.age \geq secondHighest.age$ **then**
        $b \leftarrow highest$
      **else**
        $random \leftarrow drawUniformDistribution(0, 1)$
        **if** $random < np$ **then**
          $b \leftarrow secondHighest$
        **else**
          $b \leftarrow highest$
        **end if**
      **end if**
    **end if**
    $\mathcal{A} \leftarrow \mathcal{A} \bigcup \{b\} \setminus \mathcal{N}(b)$
    **if** $\mathcal{A}.value > \mathcal{W}.value$ **then**
      $\mathcal{W} \leftarrow \mathcal{A}$
    **end if**
  **end for**
**end while**

Algorithm 1 sketches how the winner determination problem is solved. The algorithm starts initialising the set $\mathcal{B}$ containing all the bids received so far. The winner set $\mathcal{W}$ is initialised to the empty set. Once the initialisation has been concluded, the algorithm executes the main loop for 1 second. Within the main loop, a stochastic search is performed for a number of steps equal to the number of bids in $\mathcal{B}$. Set $\mathcal{A}$ contains the candidate bids for the winner set. Then, with probability $wp$ (walk probability[8]), a random bid is selected from the set of bids that are not actually in the candidate winner set ($\mathcal{B} \setminus \mathcal{A}$), while, with probability $1 - wp$, the highest and the second highest bids are evaluated. The highest bid is selected if its age (i.e., the number of steps since a bid was last selected to be added to a candidate solution) is greater than or equal to the age of the second highest bid. Otherwise, with probability $np$ (novelty probability[9]) the second highest, and with probability $1 - np$ the highest bid is selected. Finally bid $b$ is added to the candidate solution $\mathcal{A}$ and all its neighbours $\mathcal{N}(b)$, that is, the set of bids over bundles that share with $b$ at least one item, are removed from $\mathcal{A}$. Finally, if the value of $\mathcal{A}$ (i.e., the sum of the bids in $\mathcal{A}$) is greater than the value of the best-so-far winner set, $\mathcal{W}$, the best solution found so far is updated.

---
8. The probability of adding a random, not previously allocated bid to the candidate winner set.

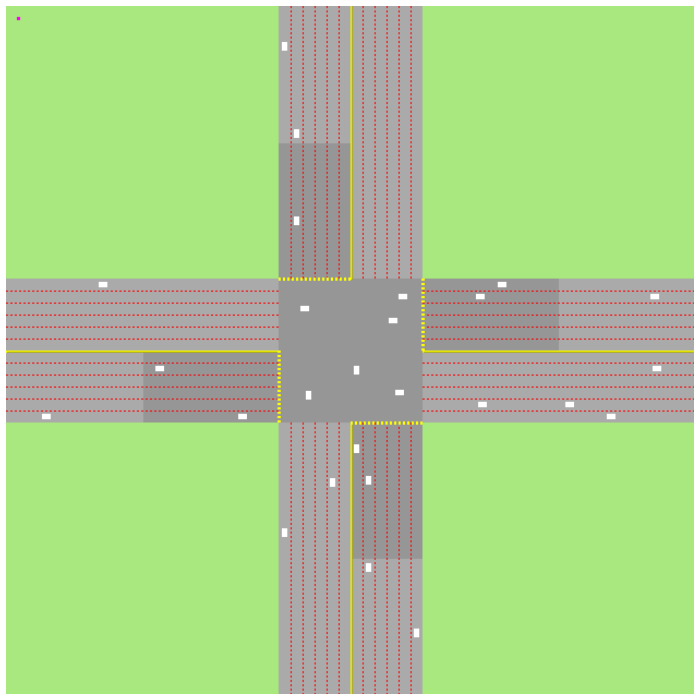

Figure 6: Simulator of a single intersection

### 4.2 Simulation Environment

The simulator we use for the evaluation of our auction-based policy is a custom, microscopic, time-and-space-discrete simulator, with simple rules for acceleration and deceleration. The simulated area is modelled as a grid, and subdivided in lanes (see Figure 6). Each lane is 3m wide, and subdivided in 12 squared tiles of 0.25m each. Each vehicle is modelled as a rectangle of $8 \times 16$ tiles, or equivalently, as a rectangle of $2m \times 4m$, and has a preferred speed in the interval $[30, 50]$km/h. The simulation environment generates the origin-destination pair randomly. When a vehicle is spawned inside the simulation, it is inserted at the beginning of one of the 4 incoming links, randomly selected, and a destination is randomly assigned to it. The destination implies the type of turn (left, right or straight) that the vehicle will perform at the intersection as well as the lane it will use to travel (the left-most lane in case of left turn, the right-most lane in case of right turn, any lane for going straight). The preferred speed is assigned using a normal distribution with mean 40km/h and variance 5km/h, while being limited by the interval $[30, 50]$.

---

9. The probability of adding to the candidate winner set the second highest bid rather then the "greedy" bid, i.e., the highest in value.

Since the link used to approach the intersection is relatively short, we assume that each vehicle will travel in its pre-assigned lane, without changing it. Therefore, we only need a car-following model to simulate the vehicle dynamics, and no lane-changing model is needed. The car-following model we use is the Intelligent Driver Model (Treiber, Hennecke, & Helbing, 2000). In this model, the decision of any driver to accelerate or to brake depends only on its own speed, and on the speed of the vehicle immediately ahead of it. Specifically, the acceleration $dv/dt$ of a given vehicle depends on its speed $v$, on the distance $s$ to the front vehicle, and on the speed difference $\Delta v$ (positive when approaching) :

$$\frac{dv}{dt} \;=\; a \cdot \left[1 - \left(\frac{v}{v_p}\right) - \left(\frac{s^*}{s}\right)^2\right] \tag{1}$$

where

$$s^* \;=\; s_0 \;+\; \left(v \cdot T \;+\; \frac{v \cdot \Delta v}{2 \cdot \sqrt{a \cdot g}}\right) \tag{2}$$

and $a$ is the acceleration, $g$ is the deceleration[10], $v$ is the actual speed, $v_p$ is the preferred speed, $s_0$ is the minimum gap, $T$ is the time headway.

The acceleration is divided into an acceleration towards the preferred speed on a free road, and braking decelerations induced by the front vehicle. The acceleration on a free road decreases from the initial acceleration $a$ to 0 when approaching the preferred speed $v_p$.

The braking term is based on a comparison between the "preferred distance" $s^*$, and the current gap $s$ with respect to the front vehicle. If the current gap is approximately equal to $s^*$, then the braking deceleration essentially compensates the free acceleration part, so the resulting acceleration is nearly zero. This means that $s^*$ corresponds to the gap when following other vehicles in steady traffic conditions. In addition, $s^*$ increases dynamically when approaching slower vehicles and decreases when the front vehicle is faster. As a consequence, the imposed deceleration increases with decreasing distance to the front vehicle, increasing its own speed, and increasing speed difference to the front vehicle. The aforementioned parameters were set to $v_p = 50\text{km/h}$, $T = 1.5\text{s}$, $s_0 = 2\text{m}$, $a = 0.3\text{m/s}^2$, $b = 3\text{m/s}^2$. The speed of a vehicle is updated every second, and its position, since the space is discrete, is updated to the tile closest to the new position in the continuous space.

### 4.3 Experimental Results

We create different traffic demands by varying the expected number of vehicles ($\lambda$) that, for every O-D pair, are spawned in an interval of 60 seconds, using a Poisson distribution. We spawned vehicles for a total time of 30 minutes. Table 1 shows the number of vehicles that have been generated for different values of $\lambda$.

The main goal of this set of experiments is to test whether the policy based on combinatorial auction (CA) enforces an inverse relation between money spent by the bidders and their delay. The delay measures the increase in travel time due to the presence of the intersection. It is computed as the difference between the travel time when the intersection is regulated by the intersection manager, and the travel time that would arise if the vehicle could travel unhindered through the intersection. The bid that a driver is willing to submit is drawn from a normal distribution with mean 100 cents and variance 25 cents, since the willingness of human drivers to pay is usually normally (or log-normally) distributed (Hensher & Sullivan, 2003). Thus, the agents are not homogeneous in the sense that the amount of money that they are offering differs from one to another. In this population, we track the delay of a subset of drivers, which are endowed with 10, 50, 100, 150, 200, 1000, 1500, 2000 and 10000 cents. This endowment is entirely allocated as a bid. We also evaluate the auction-based policy with respect to the average delay of the entire population of drivers.

---

10. $a$ and $g$ are different parameters with different values, since usually a vehicle decelerates (i.e., brakes) more strongly than it accelerates.

| $\lambda$ | 1 | 5 | 10 | 15 | 20 | 25 | 30 |
|---|---|---|---|---|---|---|---|
| # of vehicles | 29 | 136 | 285 | 438 | 633 | 716 | 832 |

Table 1: Traffic demands for a single intersection

For the WDP algorithm, we set the walk probability $wp = 0.15$ and the novelty probability $np = 0.5$, as these values produced the best results in auctions of similar type and size (Hoos & Boutilier, 2000). In all the experiments, we give the intersection manager one second to execute the WDP algorithm and return a solution. To give more time to bidders to submit their bids, before starting another auction, the intersection manager waits another second to collect incoming bids[11]. To determine if one second is enough for the winner determination algorithm to produce acceptable results, we performed the following experimental analysis. According to the results reported by Hoos and Boutilier, given an auction with 100 bids, the winner determination algorithm is able to find the optimal solution with a probability of 0.6, which tends to 1 if the algorithm is allowed to run for more than 10 seconds. This is encouraging, but in order to justify the adequacy of the stochastic algorithm for our particular problem, we need to show that, in the context of the auction-based policy for reservation-based intersection control, it produces results that are reasonably close to the optimum, despite the relatively short time (1 second in the experiments) that the algorithm has to return a solution. Given that the average number of submitted bids for a single auction is between 3 for low traffic demand ($\lambda = 1$) and 80 for high traffic demand ($\lambda = 30$), we performed several experiments to compare the solution provided by the algorithm with 1 second of run-time with the solution provided by the algorithm with 100 seconds. The solution provided by the second execution of the algorithm is assumed to be the best approximation of the optimal solution. The result was that the winner determination algorithm is able to find a solution whose value is at least 95% of the optimal solution value with a probability between 96.1% for high traffic demand ($\lambda = 30$) and 99.2% for low traffic demand ($\lambda = 1$).

Figure 7 plots (in logarithmic scale) the relation between travel time and bid value for different values of $\lambda$. All the error bars denote 95% confidence intervals. There is a sensible decrease of the delay experienced by the drivers that bid from 100 to 150 cents, which represent 49.8% of drivers whose bid is greater than the mean bid. Still, such delay reduction tends to settle for drivers that bid more than 1000 cents.

---

11. Nevertheless, the intersection manager runs a separate thread that receives incoming bids also during the WDP algorithm execution.

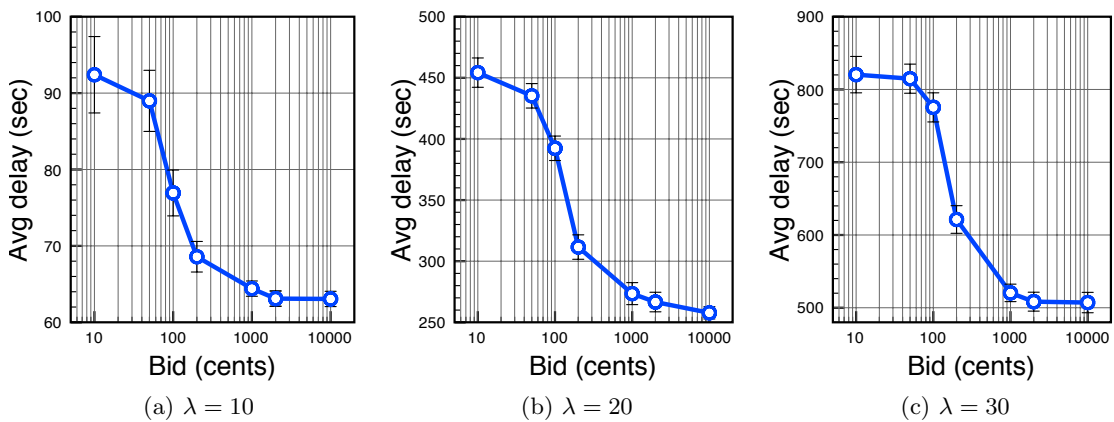

Figure 7: Bid-delay relation for various values of $\lambda$ and normally distributed endowments

We remark that the auction-based policy also uses the reservation distance as preprocessing step, which guarantees that a driver's bid cannot be rejected indefinitely. In fact, a vehicle is allowed to approach the intersection and slow down until it reaches the intersection edge. At that point, if its request is rejected because another driver submitted a higher value bid, the reservation distance is updated to the stopped vehicle's distance. Therefore, in the following time step, only this driver will be allowed to submit a bid with its preferred value. The result is, of course, that this driver will suffer greater delays compared to other drivers that are willing to pay more[12].

The auction-based allocation policy has proven to be effective regarding its main goal, that is, rewarding lower delays to those drivers that value their disputed reservations the most. However, it is worth analysing the impact that such a policy has on the intersection's average delay. Figure 8a plots the average delay for different traffic demands ($\lambda \in [1, 30]$). Again, the error bars denote 95% confidence intervals. When traffic demand is low, the performance of the CA policy and the FCFS is approximately the same. However, when traffic demand increases, there is a noticeable increase of the average delay when the intersection manager applies CA. This was somewhat expected, because the CA policy aims to grant a reservation to the driver that values it the most, rather than maximising the number of granted requests. Thus, a bid $b$, whose value is greater than the sum of $n$ bids that share some items with $b$, is likely to be selected in the winner set. If so, only 1 vehicle will be allowed to transit, while $n$ other vehicles will have to slow down and try again. This fact is highlighted also by the average rejected requests (Figure 8b). Since all the non-winning bids are rejected, the number of rejected requests with the CA policy is up to four times greater than with the FCFS policy.

---

12. Although we focus on technical problems and not social or political ones, one may wonder whether it is fair that "rich" drivers can travel faster than "poor" drivers using a road-infrastructure that is a public good. Nevertheless, we could argue that through the money raised by the auction-based policy "rich" drivers contribute much more to the maintenance and extension of the public road infrastructure than "poor" drivers.

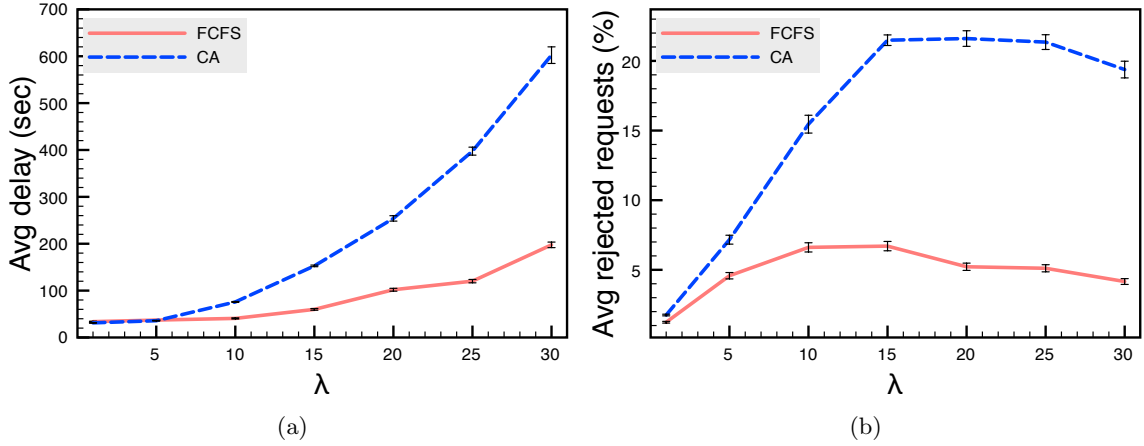

Figure 8: Average delay (a) and average rejected requests (b)

### 4.4 Discussion

The principle of optimising the use of the available resources is not the unique guiding principle of a traffic controller. In the real world, depending on the context and their personal situation, drivers value the importance of travel times and delays quite differently. Thus, it makes sense to elaborate control policies that are aware of these different valuations and that reward the drivers that value the disputed resources the most. In this respect, we evaluated a control policy for reservation-based intersections that relies on an auction mechanism. With such a policy, drivers that submit high-value bids usually experience significant reductions in their individual delays (about 30% less compared to drivers that submit low-value bids).

However, since the objective of this policy is not maximising the number of granted reservations, it pays a social cost, in the form of greater *average* travel times. This fact might limit the applicability of the CA policy in high load situations. In this case, additional mechanisms to reduce the number of vehicles that approach a single intersection are needed.

It is also worth noting how it is possible that a driver, even with a theoretically infinite amount of money, cannot experience zero delay when approaching an intersection. This is because an auction carried on in a realistic traffic scenario is quite different from a synthetic auction that has been set-up for benchmarking purposes (Hoos & Boutilier, 2000). The auctions that arise in the traffic scenario are affected by the high level of dynamism, uncertainty and noise, intrinsic to the domain. For example, in high load situations, the reservation distance plays an important role, since it filters out many potentially winning bids coming from a greater distance[13]. Figure 9 plots how the reservation distance decreases over time for different traffic demands. In high load situations, the reservation distance tends to be small, therefore a wealthy driver must reach this reservation distance in order to participate in the auction and acquire a reservation, thus increasing its travel time. The estimation of the arrival time also greatly affects the performance of the auction. In fact, in high load situations, such an estimation is much more noisy and uncertain, and it is likely that a driver must resubmit a reservation request with the updated arrival time. In this way, it is possible that an agent wins an auction at time $t$ and then, due to a new estimation of the arrival time, must resubmit its bid at time $t + \Delta t$. The bidders that participate in the auction at time $t + \Delta t$ are obviously different from those that participated at time $t$, so there is no guarantee that the agent might win the auction again.

---

13. As outlined in Section 3.2, the reservation distance is the maximum distance at which a driver is allowed to request a reservation.

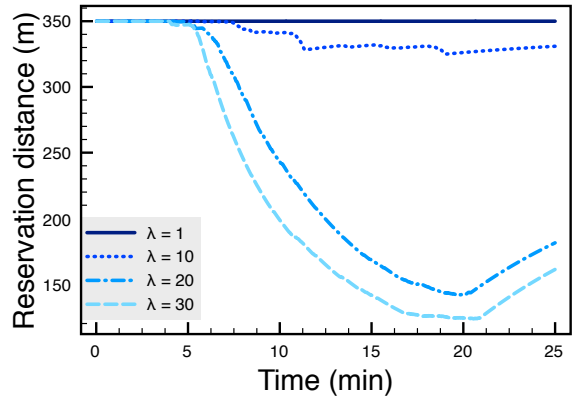

Figure 9: Reservation distance

Furthermore, a real-world scenario such as urban traffic limits the auction design space and the applicable solution methods for winner determination and payments calculation. In fact, we gave priority to the winner determination problem, adapting a local search algorithm to our needs, while for the payments calculation we did not adopt any sophisticated method, i.e., a winner pays a price that is exactly the bid that was submitted. This, as with any first-price payment mechanism, could in principle lead to malicious behaviours, with drivers that try to acquire reservations by submitting bids that are lower than the real valuations they have. In single item auctions it is computationally easy to set up an incentive compatible payment mechanism, such as the second-price (Vickrey) mechanism. Unfortunately, extending this mechanism to combinatorial auctions in not (computationally) straightforward, since the equivalent truth-revealing mechanism in the combinatorial world, the Vickrey-Clarke-Groves (VCG) payment mechanism (Clarke, 1971; Groves, 1973; Vickrey, 1961), is NP-complete. Therefore, although a driver agent could potentially acquire a reservation by submitting a bid $\widehat{b}$ that is lower than its real valuation $b$, from a practical point of view this exclusively affects the revenues that the auctioneer should gain if every bidder were truth-telling, which is not our primary concern. Another possible weakness is the fact that a bidder could start bidding lower than their real valuation and then raising their bid if they are not able to acquire it, thus leading to a communication overhead between bidders and auctioneer. Nevertheless, only the bidders within the reservation distance are able to submit a bid, thus the number of bids that the intersection manager may receive simultaneously is necessarily bounded.

## 5. Network of Intersections

In the single intersection scenario we analysed the performance of an auction-based policy for the allocation of reservations. In that context, the driver was modelled as a simple agent that selects the preferred value for the bid that will be submitted to the auctioneer. If we focus on an urban road network with multiple intersections, it is interesting to notice that the decision space of a driver is much broader. In fact, drivers are involved in complex and mutually dependent decisions such as route choice and departure time selection. At the same time, this scenario opens new possibilities for intersection managers to affect the behaviour of drivers. For example, an intersection manager may be interested in influencing the collective route choice performed by the drivers, using variable message signs, information broadcast, or individual route guidance systems, so as to evenly distribute the traffic over the network. This problem is called traffic assignment.

In Section 5.1 we evaluate how market-inspired methods (Gerding et al., 2010) can be applied as traffic assignment strategies for networks of reservation-based intersections. The idea is that, if there is a market where drivers acquire the necessary reservations to pass through the intersections of the urban network, this market, and the intersection managers that operate in it from the supply side, can be designed to work as a traffic assignment system. In particular, we model the intersection managers so that they apply a competitive pricing strategy to compete among themselves for the supply of the reservations that are traded. Finally, in Section 5.2 we combine this traffic assignment strategy with the auction-based control policy into an integrated mechanism for traffic management of urban road networks.

### 5.1 Competitive Traffic Assignment (CTA)

Traffic assignment strategies aim at influencing the collective route choice of drivers in order to use the road network capacity efficiently. Therefore, we can see the traffic assignment problem as a distributed choice and allocation problem, since a set of resources (i.e., the links capacity) must be allocated to a set of agents (i.e., the drivers). To this regard, markets as mediators for distributed resource allocation problems have been applied to several socio-technical systems (Gerding et al., 2010).

Setting out from the approach outlined in the work by Vasirani and Ossowski (2011), we follow this metaphor and model each intersection manager as a provider of the resources, in this case, the reservations of the intersection it manages. Thus, each intersection manager is free to establish a price for the reservations it provides. On the other side of the market, each driver is modelled as a buyer of these resources. Provided with the current prices of the reservations, it chooses the route, according to its personal preferences about travel times and monetary costs. Each intersection manager is modelled so as to compete with all others for the supply of the reservations that are traded. Therefore, our goal as market designers is making the intersection managers adapt their prices towards a price vector that accounts for an efficient allocation of the resources.

### 5.1.1 CTA Pricing Strategy

Let $\mathcal{L}$ be the set of incoming links of a generic intersection. For each incoming link $l \in \mathcal{L}$, the intersection manager defines the following variables:

- Current price $p^t(l)$: is the price applied by the intersection manager to the reservations sold to the drivers that come from the incoming link $l$.

- Total demand $d^t(l \mid p^t(l))$: represents the total demand of reservations from the incoming link $l$ that the intersection manager observes at time $t$, given the current price $p^t(l)$. It is given by the number of vehicles that want to cross the intersection coming from link $l$ at time $t$.

- Supply $s(l)$: defines the reservations supplied by the intersection manager for the incoming link $l$. It is a constant and represents the number of vehicles that cross the intersection coming from link $l$ that the intersection manager is willing to serve.

- Excess demand $z^t(l \mid p^t(l))$: is the difference between the total demand at time $t$ and the supply, $z^t(l \mid p^t(l)) = d^t(l \mid p^t(l)) - s(l)$.

Given the set of all the intersection managers that are operating in the market, $\mathcal{J}$, we define the price vector $\mathbf{p}^t$ as the vector of the prices applied by each intersection manager to each of its controlled links:

$$\mathbf{p}^t = [\; p_1^t(l^1) \; p_1^t(l^2) \; \ldots \; p_{|\mathcal{J}|}^t(l^h) \;] \qquad (3)$$

where $p_1(l^1)$ is the price applied by intersection manager 1 to its controlled link $l^1$, $p_1(l^2)$ is the price applied by the same intersection manager to another link $l^2$ of its intersection, and $p_{|\mathcal{J}|}(l^h)$ is the price applied by the $|\mathcal{J}|$th intersection manager to its last controlled link $l^h$.

In particular, we say that a price vector $\mathbf{p}^t$ maps the supply with the demand if the excess demand $z^t(l \mid p^t(l))$ is 0 for all links of the network. This price vector, which corresponds to the market equilibrium price, can be computed through a Walrasian auction (Codenotti, Pemmaraju, & Varadarajan, 2004), where each buyer (i.e., driver) communicates to the suppliers (i.e., intersection managers) the route that it is willing to choose, given the current price vector $\mathbf{p}^t$. With this information, each intersection manager computes the demand $d^t(l \mid p^t(l))$ as well as the excess demand $z^t(l \mid p^t(l))$ for each of its controlled links. Then, each intersection manager adjusts the prices $p^t(l)$ for all the incoming links, lowering them if there is excess supply ( $z^t(l \mid p^t(l)) < 0$ ) and raising them if there is excess demand ( $z^t(l \mid p^t(l)) > 0$ ). The new price vector $\mathbf{p}^{t+1}$ is communicated to the drivers that iteratively choose their new desired route, on the basis of the new price vector $\mathbf{p}^{t+1}$. Once the equilibrium price is computed, the trading transactions take place and each driver buys the required reservations at the intersections that lay on its route.

The Walrasian auction relies on quite strict assumptions, which make a direct implementation in the traffic domain hard. For instance, the set of buyers is assumed to be fixed during the auction, which means for the traffic domain that new drivers may not join an auction until it terminates. Also the fact that no transactions can take place at disequilibrium prices is a strict assumption for the traffic domain. It is unreasonable for all the

---

**Algorithm 2** Intersection manager price update
---
$t \leftarrow 0$
**for all** $l \in \mathcal{L}$ **do**
  $p^t(l) \leftarrow \delta$
  $s(l) \leftarrow 0.5 \cdot \mu_{\text{opt}} \cdot \ell(l)$
**end for**
**while** true **do**
  **for all** $l \in \mathcal{L}$ **do**
    $d^t(l) \leftarrow evaluateDemand$
    $z^t(l) \leftarrow d^t(l) - s(l)$
    $p^t(l) \leftarrow p^t(l) + p^t(l) \cdot \dfrac{z^t(l)}{s(l)}$
  **end for**
  $t \leftarrow t + 1$
**end while**

---

drivers to wait to reach the equilibrium point before choosing the desired route and starting to travel. Finally, a driver is probably willing to transfer money to an intersection manager when it is spatially close to it, that is, when it is already travelling along its desired route.

Thus, we implement a pricing strategy that aims to reach the equilibrium price - as in the Walrasian auction - but that works on a continuous basis, with drivers that leave and join the market dynamically, and with transactions that take place continuously. To reach general equilibrium, each intersection manager applies the price update strategy sketched in Algorithm 2. At time $t$, each intersection manager independently computes the excess demand $z^t(l \mid p^t(l))$ and updates the price $p^t(l)$ using the formula (Codenotti et al., 2004):

$$p^{t+1}(l) \leftarrow \max \left[ \delta, \ p^t(l) + p^t(l) \cdot \frac{z^t(l \mid p^t(l))}{s(l)} \right] \quad (4)$$

where

- $\delta$ is the minimum price that an intersection manager charges for the reservations that it sells.

- $s(l)$ is the supply of the intersection manager, that is, the number of vehicles above which the intersection manager considers there is excess demand and starts to raise prices.

We claim that drivers that travel through road network links with low demand shall not incur any costs. For this reason, we choose $\delta = 0$. To define the supply $s(l)$, we rely on the fundamental diagram of traffic flow (Gerlough & Huber, 1975). Let $\mu_{\text{opt}}$ be the density that maximises the traffic flow on link $l$ (see Figure 10). We choose $s(l) = 0.5 \cdot \mu_{\text{opt}} \cdot \ell(l)$, where $\ell(l)$ is the length of link $l$. In other words, the intersection manager considers that there is excess demand when the density reaches 50% of optimal density. In this way the intersection manager aims to avoid exceeding $\mu_{\text{opt}}$ by raising prices and diverting drivers to different routes before reaching $\mu_{\text{opt}}$.

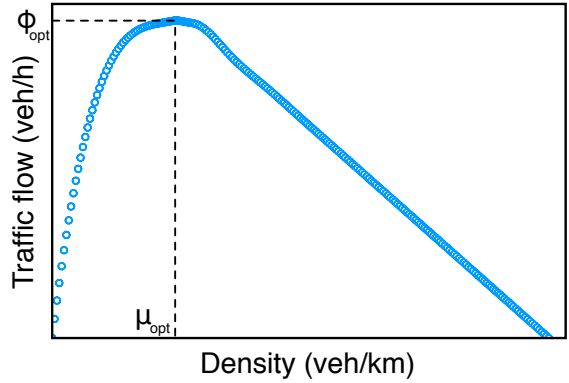

Figure 10: Fundamental diagram of traffic flow

### 5.1.2 Driver Model

Unlike the single intersection scenario, in this case we need a reasonable driver model for the route choice. The route choice problem is modelled as a multi-attribute utility-function-maximisation problem. Given that the traffic system is regulated by a market mechanism, the driver must take into consideration different aspects of a route to determine its utility value. A route $\rho$ is modelled as an ordered list of links, $\rho = [l^1 \ldots l^N]$. A generic link $l^k$ is characterised by two attributes: the estimated travel time $\mathbb{E}[T(l^k)]$ and the price of reservations $K(l^k)$. For sake of simplicity, the estimation is based on the travel time at free flow, and does not consider real-time information of traffic conditions (see Equation 5, where $\ell(l^k)$ is the length of link $l^k$, and $v_{\max}(l^k)$ is the maximum allowed speed on link $l^k$). The price of reservations of link $l^k$ is always 0, unless the link $l^k$ is one of the incoming link of an intersection ($l^k = l$), in which case the price is $p^t(l)$ (Equation 6).

$$\mathbb{E}[T(l^k)] = \frac{\ell(l^k)}{v_{\max}(l^k)} \tag{5}$$

$$K(l^k) = \begin{cases} p^t(l) & \text{if } l^k = l \in \mathcal{L} \\ 0 & \text{otherwise} \end{cases} \tag{6}$$

The summatory of the estimated travel time over all the links of $\rho$ gives the estimated travel time of the entire route $\rho$:

$$\mathbb{E}[T(\rho)] = \sum_{k=1}^{N} \mathbb{E}[T(l^k)] \tag{7}$$

Similarly, the summatory of the price of reservations over all the links of $\rho$ gives the price of the entire route $\rho$:

$$K(\rho) = \sum_{k=1}^{N} K(l^k) \tag{8}$$

Let $\mathcal{C} = \{\rho_1, \ldots, \rho_M\}$ be the choice set, that is, the set of routes available to a driver. The set $\mathcal{C}$ is built using a $k$-shortest paths algorithm (Yen, 1971), with $k = 10$. Let $u_T(\rho)$ be the normalised utility of route $\rho$ against the estimated travel time attribute (Equation 9), where $M_T = \max\limits_{\rho_i \in \mathcal{C}} \mathbb{E}[T(\rho_i)]$ and $m_T = \min\limits_{\rho_i \in \mathcal{C}} \mathbb{E}[T(\rho_i)]$.

$$u_T(\rho) = \frac{M_T - \mathbb{E}[T(\rho)]}{M_T - m_T} \tag{9}$$

Let $u_K(\rho)$ be the normalised utility of route $\rho$ against the reservations cost attribute (Equation 10), where $M_K = \max\limits_{\rho_i \in \mathcal{C}} K(\rho_i)$ and $m_K = \min\limits_{\rho_i \in \mathcal{C}} K(\rho_i)$.

$$u_K(\rho) = \frac{M_K - K(\rho)}{M_K - m_K} \tag{10}$$

The driver multi-attribute utility of route $\rho$ is then defined as:

$$U(\rho) = w_T \cdot u_T(\rho) + w_K \cdot u_K(\rho) \tag{11}$$

where $w_T$ is the weight of the estimated travel time attribute and $w_K$ is the weight of the cost of reservations attribute. Basically, if $w_T = 1$ the driver utility only considers the attribute related to the estimated travel time (i.e., it prefers the shortest route, no matter the price of the reservations), if $w_K = 1$ the driver utility only considers the attribute related to the cost of reservations (i.e., it prefers the cheapest route, no matter the travel time), while for every other combination of the weights $w_T$ and $w_K$ the driver considers the trade-off between estimated travel time and cost of reservations. In the experiments we draw $w_T$ from a uniform distribution over the interval $[0, 1]$, and we set $w_K = 1 - w_T$.

Once the utility of the routes that form the choice set $\mathcal{C}$ has been computed, the driver must choose one of these alternatives. In this work, we model the driver as a deterministic utility maximiser that always selects the route with the highest utility value. Since the price of the incoming links of an intersection is changing dynamically, the term $u_K(\rho)$ in Eq. 11 may change during the journey. For this reason, the driver continuously evaluates the utility of the route it is following and, in case that a different route becomes more attractive, it may react and change on-the-fly how to reach its destination, selecting a route different from the original one.

### 5.1.3 SIMULATION ENVIRONMENT

The experimental evaluation is performed on a hybrid mesoscopic-microscopic simulator, where the traffic flow on the roads is modelled at mesoscopic level (Schwerdtfeger, 1984), while the traffic flow inside the intersections is modelled at microscopic level (Nagel & Schreckenberg, 1992).

In a mesoscopic model vehicle dynamics is governed by the average traffic density on the link it traverses rather than the behaviour of other vehicles in the immediate neighbourhood as in microscopic models. A road network is modelled as a graph, where the nodes represent intersections and the edges represent the lanes of a road. An edge, also called stretch, is subdivided into sections (of typically 500m length) for which a constant traffic condition is assumed. A vehicle $i$ that at time $t$ is driving on a link $l^k$ is characterised by its position $x_i^t \in [0, \ell(l^k)]$, and its speed $v_i^t$. At each time step, a new target speed for each vehicle is computed, using the formula:

$$\widehat{v}_i^{t+\Delta t} = (1 - \frac{x_i^t}{\ell(l^k)}) \cdot y(l^k) + \frac{x_i^t}{\ell(l^k)} \cdot y(l^{k+1}) \tag{12}$$

where $y(l^k)$ is the reference speed of link $l^k$ and $y(l^{k+1})$ is the reference speed of link $l^{k+1}$. Such reference speeds are calculated by taking into consideration the mean speed of the link and the vehicle's desired speed. The mean speed of the link is calculated with a speed-density function that for a given link's density $\mu(l^k)$ returns the link's mean speed (Schwerdtfeger, 1984).

The equation above takes into consideration the fact that the closer the vehicle is to the next link $l^{k+1}$, the higher is the effect of the link reference speed on the vehicle target speed. If the new target speed $\widehat{v}_i^{t+\Delta t}$ is higher (lower) than the current speed $v_i^t$, the vehicle accelerates (decelerates) with a vehicle-type specific maximum acceleration (deceleration). The new speed is then denoted by $v_i^{t+\Delta t}$. Finally, the vehicle position is updated using the formula:

$$x_i^{t+\Delta t} = x_i^t + \frac{1}{2} \cdot (v_i^t + v_i^{t+\Delta t}) \cdot \Delta t \tag{13}$$

If $x_i^{t+\Delta t} \geq \ell(l^k)$, the vehicle is placed in the next link of its route, the densities for link $l^k$ and $l^{k+1}$ are updated accordingly, and the position is reset to $x_i^{t+\Delta t} - \ell(l^k)$.

The mesoscopic model described above does not offer the necessary level of detail to model a reservation-based intersection. For this reason, when a vehicle enters an intersection, its dynamics switches into a microscopic, cellular-based, simulator (Nagel & Schreckenberg, 1992), similar to the simulation environment used in Section 4.2. Still, the cells that compose the intersection's area are more coarse grained (5 meters), and for simplicity we assume that the vehicles cross the intersection at a constant speed, so that any additional tuning of parameters, such as slowdown probability or acceleration/deceleration factors, is not necessary.

### 5.1.4 Experimental Results

Although our work does not depend on the underlying road network, we chose a (simplified) topology of the entire urban road network of the city of Madrid for our empirical evaluation (see Figure 11). The network is characterised by several freeways that connect the city centre with the surroundings and a ring road. Each large dark vertex in Figure 11 - if it connects three or more links - is modelled as a reservation-based intersection. We aim to recreate a typical high load situation (i.e., the central, worst part of a morning peak), with more than 11,000 vehicles departing within a time window of 50 minutes (see Table 2). The vehicles that travel to and from 7 destinations outside the city (marked with $O_1$ up to $O_7$ in Figure 11) form the traffic under evaluation.

The market-inspired traffic assignment strategy is compared with a network of FCFS reservation-based intersections. In the latter, the drivers' route choice only takes into consideration the expected travel time at free flow, since there is no notion of price.

We focus on two different types of metrics, one related to the vehicles and one related to the network. The network-related metric is the density variation over time at 7 critical intersections (marked with $c_1 \ldots c_7$ in Figure 11), which connect the freeways going toward the city centre with the ring road. The vehicle-related metric is the average travel time, grouped by the origin-destination (O-D) pair. For a given O-D pair, we compute the average travel time of the vehicles that go from O to D. This measurement is then averaged over 30 runs. Furthermore, for each O-D pair we compute the improvement $\%\Delta$ of CTA over FCFS based on the average travel times. Table 3 shows the average travel time of the drivers, according to their origin-destination pairs, when the reservations are allocated through the competitive traffic assignment (CTA) and when they are granted with the usual FCFS policy. Using CTA we observe a net reduction of the average travel time for 30 of 42 origin-destination pairs. Such reduction is generally noteworthy for the busiest[14] routes, such as $O_6$-$O_2$, $O_6$-$O_3$ and $O_7$-$O_3$. Along the some of the less demanded O-D pairs, FCFS is the best performing policy. This happens when on the most preferred route from O to D the traffic density is already low enough to assure free flow, but there exist alternative routes with even lower demand, and CTA keeps diverting traffic along these potentially longer and thus slower routes.

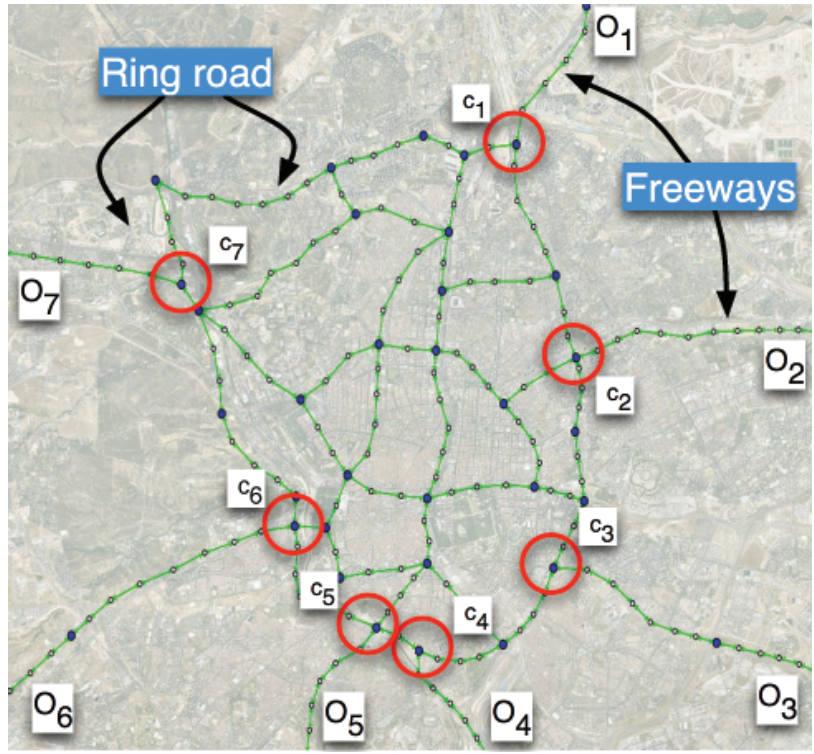

Figure 11: Urban road network

|  | Destination | | | | | | |
|---|---|---|---|---|---|---|---|
| **Origin** | $O_1$ | $O_2$ | $O_3$ | $O_4$ | $O_5$ | $O_6$ | $O_7$ |
| $O_1$ | - | 323 | 355 | 336 | 311 | 349 | 271 |
| $O_2$ | 223 | - | 221 | 248 | 191 | 214 | 229 |
| $O_3$ | 300 | 364 | - | 343 | 358 | 368 | 362 |
| $O_4$ | 208 | 233 | 229 | - | 218 | 199 | 204 |
| $O_5$ | 199 | 228 | 261 | 216 | - | 238 | 209 |
| $O_6$ | 290 | 316 | 398 | 386 | 374 | - | 337 |
| $O_7$ | 224 | 231 | 214 | 235 | 219 | 253 | - |

Table 2: OD Matrix (# of vehicles)

To evaluate the effects of the trading activity between drivers and intersection managers it is worth observing the density variation over time at the critical intersections $c_1$ to $c_7$, plotted in Figure 12. In general, density tends to be lower with CTA compared with the system regulated by FCFS intersection managers. At the least demanded intersections $c_1$, $c_2$ and $c_7$, that is, those intersections whose density is below the density that maximises traffic flow (see Figure 10), there is no substantial difference between CTA and FCFS. These critical intersections are less demanded due to the topology of the network. In fact, fewer origins are located in the northern part ($O_1$, $O_2$ and $O_7$).

At the critical intersections $c_3$, $c_4$ and $c_6$, the vehicle density with CTA is always below the density that results from the use of FCFS, especially in the case of intersections $c_4$ and $c_6$ where with CTA the density exceeds the optimal one by only a small extent and for a limited period of time.

At intersection $c_5$, the density has a higher peak around 9:30, but the density starts to exceed the optimal density later and begins to fall below the optimal density earlier. We calculated the integral of the density curves, measured in the interval when the curve is above the optimal density (Eq. 14)

$$\int_{t_1}^{t_2} \mu_{\text{CTA}}(t)dt \quad \text{and} \quad \int_{t_1}^{t_2} \mu_{\text{FCFS}}(t)dt \qquad (14)$$

where $\mu_{\text{CTA}}$ and $\mu_{\text{FCFS}}$ are the density functions, $t_1 = \min(\, t \mid \mu_{\text{CTA}}(t) > \mu_{\text{opt}}, \ t \mid \mu_{\text{FCFS}}(t) > \mu_{\text{opt}}\, )$ and $t_2 = \max(\, t \mid \mu_{\text{CTA}}(t) < \mu_{\text{opt}}, \ t \mid \mu_{\text{FCFS}}(t) > \mu_{\text{opt}}\, )$. This metric is lower when the reservations are allocated through the competitive market (70.24 veh·h/km versus 105.07 veh·h/km).

The result of the application of the market-inspired traffic assignment strategy is a more balanced urban network, since the price fluctuations force demand to change towards less expensive intersections. Such fluctuations contribute to creating a system in dynamic equilibrium, where unused intersections became cheaper while congested ones became more expensive. The effect is that average travel time decreases, although there are no guarantees that those drivers that pay more are rewarded with lower travel times.

---

14. We empirically noticed in the experiments that the southern part of the network tends to be more congested during the simulation. This is due to the fact that 4 of 7 origins/destinations ($O_3$, $O_4$, $O_5$, $O_6$) are located in the southern part.

|  |  | \multicolumn{7}{c}{**Destination**} |
|---|---|---|---|---|---|---|---|
|  |  | $O_1$ | $O_2$ | $O_3$ | $O_4$ | $O_5$ | $O_6$ | $O_7$ |
| **Origin** | | | | | | | | |
| $O_1$ | CTA | - | 12.09 ± 0.27 | 19.58 ± 0.80 | 26.70 ± 1.04 | 30.75 ± 0.83 | 21.17 ± 0.20 | 14.13 ± 0.12 |
|  | FCFS | - | 11.98 ± 0.31 | 22.89 ± 1.17 | 35.13 ± 1.80 | 43.57 ± 1.89 | 21.35 ± 0.40 | 13.83 ± 0.09 |
|  | %Δ | - | -0.8% | 14.4% | 24.0% | 29.4% | 0.8% | -2.2% |
| $O_2$ | CTA | 11.26 ± 0.17 | - | 14.17 ± 0.72 | 19.02 ± 0.66 | 23.72 ± 0.83 | 24.00 ± 0.40 | 20.88 ± 0.23 |
|  | FCFS | 10.15 ± 0.06 | - | 16.50 ± 1.06 | 25.87 ± 1.51 | 31.05 ± 2.03 | 38.09 ± 1.82 | 19.51 ± 0.15 |
|  | %Δ | -11.0% | - | 14.1% | 26.5% | 23.6% | 37.0% | -7.0% |
| $O_3$ | CTA | 15.57 ± 0.33 | 10.79 ± 0.14 | - | 9.18 ± 0.08 | 13.99 ± 0.37 | 18.54 ± 0.32 | 24.95 ± 0.42 |
|  | FCFS | 13.35 ± 0.09 | 9.76 ± 0.03 | - | 12.21 ± 0.62 | 17.64 ± 0.92 | 23.69 ± 6.34 | 31.73 ± 1.36 |
|  | %Δ | -16.7% | -10.6% | - | 24.8% | 20.7% | 21.7% | 21.4% |
| $O_4$ | CTA | 24.79 ± 0.77 | 20.39 ± 0.60 | 11.62 ± 0.41 | - | 8.21 ± 0.27 | 14.35 ± 0.48 | 21.66 ± 0.75 |
|  | FCFS | 26.94 ± 1.31 | 22.58 ± 1.06 | 13.92 ± 0.82 | - | 10.05 ± 0.48 | 15.74 ± 0.73 | 22.74 ± 0.99 |
|  | %Δ | 8.0% | 9.7% | 16.5% | - | 18.3% | 8.8% | 4.8% |
| $O_5$ | CTA | 26.80 ± 0.84 | 22.83 ± 0.71 | 16.30 ± 0.67 | 7.47 ± 0.20 | - | 11.11 ± 0.24 | 19.47 ± 0.63 |
|  | FCFS | 32.17 ± 1.83 | 30.61 ± 1.70 | 21.54 ± 1.39 | 8.83 ± 0.31 | - | 10.77 ± 0.26 | 17.66 ± 0.52 |
|  | %Δ | 16.7% | 25.4% | 24.3% | 15.4% | - | -3.1% | -10.3% |
| $O_6$ | CTA | 23.17 ± 0.50 | 27.31 ± 0.55 | 25.30 ± 0.89 | 16.40 ± 0.73 | 12.12 ± 0.46 | - | 16.58 ± 0.89 |
|  | FCFS | 22.51 ± 0.40 | 57.01 ± 3.13 | 41.05 ± 2.59 | 24.68 ± 1.62 | 19.02 ± 1.50 | - | 13.73 ± 0.32 |
|  | %Δ | -2.9% | 52.1% | 38.4% | 33.6% | 36.3% | - | -20.8% |
| $O_7$ | CTA | 15.05 ± 0.22 | 23.52 ± 0.33 | 31.67 ± 0.82 | 24.44 ± 0.97 | 19.12 ± 0.69 | 11.69 ± 0.11 | - |
|  | FCFS | 14.31 ± 0.10 | 23.26 ± 0.40 | 56.42 ± 3.01 | 34.99 ± 2.15 | 31.24 ± 2.06 | 12.00 ± 0.20 | - |
|  | %Δ | -5.2% | -1.1% | 43.9% | 30.2% | 38.8% | 2.5% | - |

Table 3: Average travel time in minutes (± 95%CI): CTA vs. FCFS

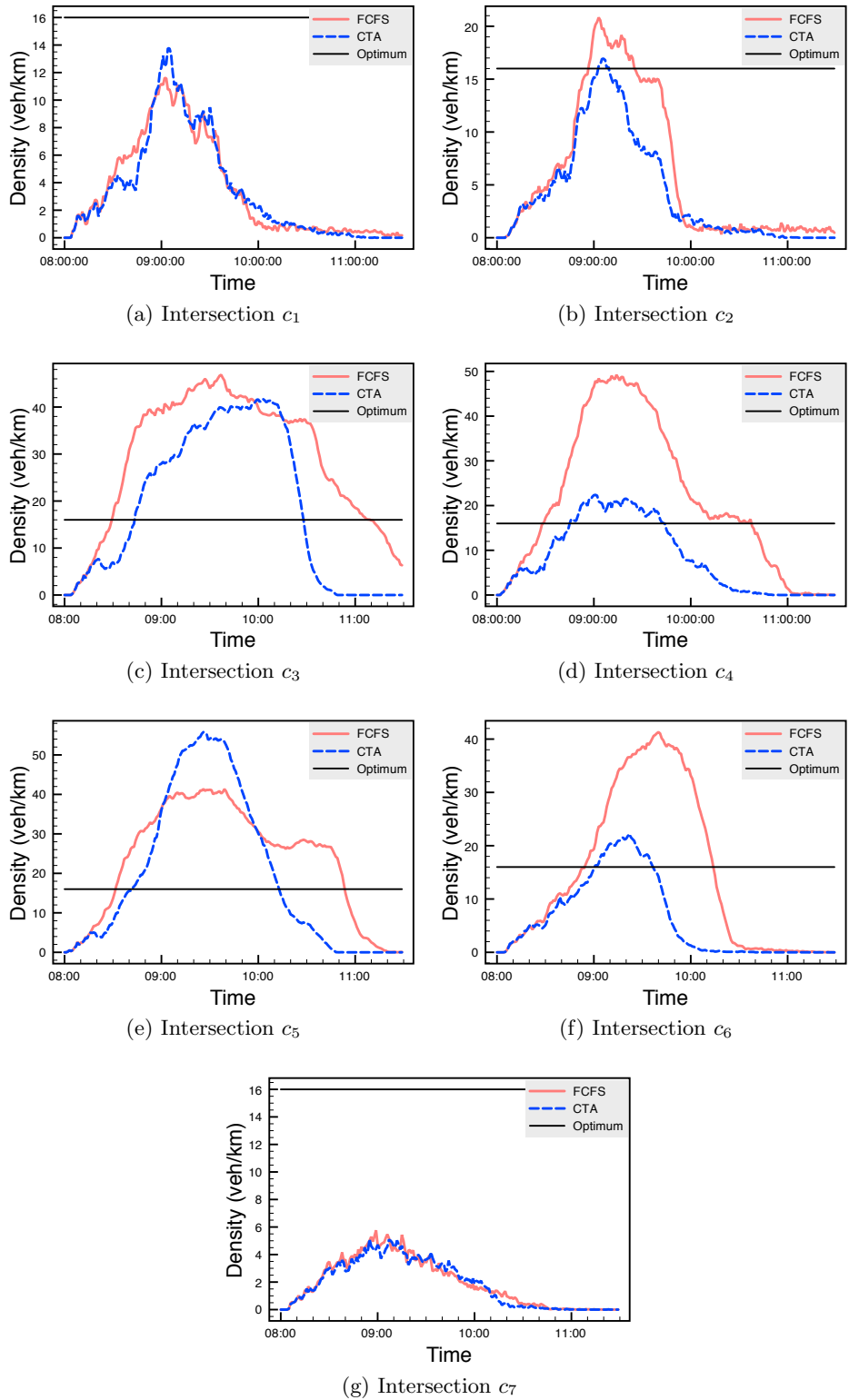

Figure 12: Density variation over time at the critical intersections under evaluation

### 5.2 An Integrated Mechanism for Traffic Management (CA-CTA)

In Section 4.1, we introduced an auction-based policy for the control of a single intersection. The experimental results showed that this policy was quite effective in allocating the reservations to the drivers that value them the most. Drivers that bid high usually experience a great reduction in delay (about 30%), compared to those drivers that submit low-value bids. However, this policy on its own showed a couple of drawbacks. First, it fosters the attainment of a user optimum rather than a global one. It therefore pays a social price, in the form of greater average delay for the entire population of drivers. Furthermore, it is possible that even wealthy drivers, in high-load situations, could not get a reservation, for example due to the decreasing reservation distance.

On the other hand, one of the results of the experimental evaluation of Section 5.1 was that a traffic assignment strategy can make the task of traffic controllers easier, enforcing a better distribution of traffic demand. Therefore, it seems reasonable to combine the auction-based policy with the competitive traffic assignment strategy into an integrated, market-inspired, mechanism for traffic management.

#### 5.2.1 CA-CTA MECHANISM

We adapt the competitive traffic assignment strategy (CTA) to combine it with the auction-based policy (CA) into an integrated mechanism for traffic management (CA-CTA). Since the intersection manager is the supplier of the reservations that are allocated through the combinatorial auction, it may control the reserve price of the auctioned reservations, i.e., the minimum price at which the intersection manager is willing to sell. We model the intersection managers in such a way that they compete for the provision of reservations to the drivers, raising the reserve price in case of increasing demand or lowering it in case of decreasing demand. The reservations are allocated through the CA policy defined in Section 4.1. However, only bids whose value is above the reserve price are accepted in the bid set.

For each incoming link $l$ of a generic intersection, the intersection manager independently computes the excess demand $z^t(l \mid p_r^t(l))$ and updates the reserve price $p_r^t(l)$ using the formula:

$$p_r^{t+1}(l) \leftarrow \max\left[\,\delta_r,\ p_r^t(l) + p_r^t(l) \cdot \frac{z^t(l \mid p_r^t(l))}{s(l)}\,\right] \qquad (15)$$

where $\delta_r$ is the minimum reserve price, and $s(l)$ is the number of vehicles that the intersection manager is willing to serve. As in Section 5.1, we choose $\delta_r = 0$ and $s(l) = 0.5 \cdot \mu_{\text{opt}} \cdot \ell(l)$, where $\ell(l)$ is the length of link $l$, and $\mu_{\text{opt}}$ is the density that maximises the traffic flow on link $l$ (see Figure 10).

#### 5.2.2 DRIVER MODEL

To empirically evaluate CA-CTA we need to define a driver route choice model that takes into consideration the fact that reservations are now allocated through a combinatorial auction with a reserve price. We assume that each driver holds a private valuation of the bids that it is willing to submit to pass through the intersections of its chosen route, defined by the variable $b$. Given the monetary constraint, the driver selects the most preferred route $\rho$, taking into consideration the estimated travel time associated with the route. A route $\rho$ is modelled again as an ordered list of links, $\rho = [l^1 \ \ldots \ l^N]$, each of them characterised by two attributes, namely estimated travel time and reserve price.

The travel time estimation is based, as before, on the travel time at free flow (Equation 5). The reserve price of a link is defined as:

$$K(l^k) = \begin{cases} p_r^t(l) & \text{if } l^k = l \in \mathcal{L} \\ 0 & \text{otherwise} \end{cases} \quad (16)$$

The price of link $l^k$ is always 0, unless the link $l^k$ is one of the incoming link of an intersection ($l^k = l$), in which case the price is equal to the reserve price $p_r^t(l)$ established by the intersection manager. The summatory of the travel time over all the links of $\rho$ gives the estimated travel time at free flow of the entire route $\rho$:

$$\mathbb{E}[T(\rho)] = \sum_{k=1}^{N} \mathbb{E}[T(l^k)] \quad (17)$$

Given $b$, the driver builds the choice-set $\mathcal{C}$ as the set of the routes whose intersections have a reserve price lower than the desired bid $b$:

$$\mathcal{C} = \left\{ \rho_1, \ldots, \rho_M \mid K(l^k) \leq b \ \forall l^k \in \rho_i \right\}$$

Once the choice-set is built, the driver selects the shortest route $\rho = \operatorname*{argmin}_{\rho_i \in \mathcal{C}} \mathbb{E}[T(\rho_i)]$.

### 5.2.3 Experimental Results

We again recreate a typical high load situation, using the same network topology and OD matrix of Figure 11 and Table 2. We are interested in two different types of properties. From one side we must evaluate whether or not the integrated management mechanism (traffic control+traffic assignment) guarantees lower delays to the drivers that submit higher bids (user optimum). For this purpose, we calculate the average (percentage) increase of the travel times $D$, calculated according to Equation 18, where $T(\rho_i)$ is the observed travel time for vehicle $i$ from its origin to its destination along route $\rho_i$, and $m_T$ is the travel time from the same origin to the same destination along the shortest route if the vehicle could cross each intersection unhindered[15]. For simplicity, we refer to the percentage increase of the travel time as *normalised delay*.

$$D = \frac{T(\rho_i) - m_T}{m_T} \quad (18)$$

On the other hand, we would like to set up a system that is fair to the entire population of drivers, guaranteeing lower average delays (global optimum). Thus, we compare our integrated mechanism with a network of intersections governed by intersection managers that apply the FCFS control policy. We assume that in this case the drivers choose the shortest route from their origin to their destination, since there are no other incentives to diverge from that route. The aim is to evaluate the global performance (in terms of average travel time) of our integrated mechanism compared to the straightforward application of the FCFS policy to a network of intersections, and to detect any potential social cost similar to that reported in Section 4.3. The metrics we use to assess the performance are the *average delay* for every O-D pair, and the *moving average* of the travel time. The latter is intended to measure how the average travel time evolves during the simulation. This metric is initialised to 0 and calculated as follows: once a driver $i$ concludes its trip, the travel time $T(\rho_i)$ is computed and the moving average travel time $\overline{T}$ is updated according to Equation 19, where $n$ is the number of drivers that have completed their trips so far.

---

15. This ratio enables us to aggregate the results of drivers even though they have different origins and/or destinations.

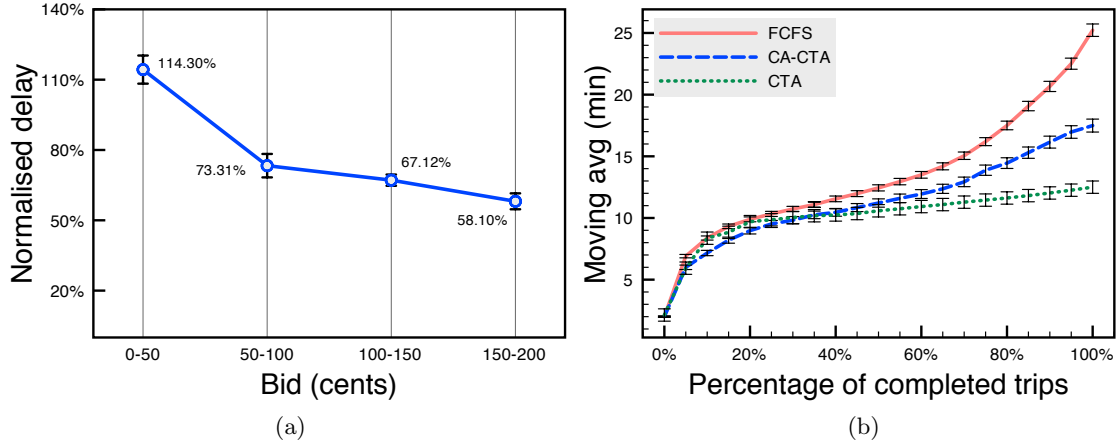

Figure 13: Relation between normalised delay and bid (a) and moving average of travel time (b)

$$\overline{T} = \overline{T} + \frac{T(\rho_i) - \overline{T}}{n+1} \qquad (19)$$

In the following tables and figures we refer to the two configurations with the abbreviations CA-CTA (which stands for "combinatorial auction-competitive traffic assignment") and FCFS.

Figure 13a plots the relation between bid value and normalised delay of the population of drivers[16]. It is still possible to appreciate an inverse relation between these two quantities: the drivers that submit bids between 150 and 200 cents reduce the delay by about 50% compared to those which bid less than 50 cents. Also at the network level, granting reservations with a combinatorial auction (the "CA" component of the CA-CTA policy) ensures that those drivers that submit higher bids experience lower delays (user optimum).

To assess the social cost incurred by CA-CTA at the global level, we measure the moving average of the travel time, that is, how the average travel time of the entire population of drivers, computed over all the O-D pairs, evolves during the simulation. We compare CA-CTA with FCFS and, for completeness, with CTA[17]. The results, with 95% confidence interval error bars, are plotted in Figure 13b. In the beginning, the average travel time is similar for all the scenarios, but as the number of drivers that populate the network (i.e., its load) increases, it grows significantly faster with FCFS than with the CA-CTA policy. In terms of average travel times CTA is the best performing policy. CA-CTA has a slightly inferior performance, but it does enforce an inverse relationship between bid value and delay (see Figure 13a). The fact that both CA-CTA and CTA outperforms FCFS is an indication that, in general, a traffic assignment strategy (the "CTA" component of both policies) improves travel time. In fact, with FCFS drivers always select the shortest route, which in some cases is not the best route choice. Furthermore, granting reservations through an auction (the "CA" component of the CA-CTA policy) ensures that bid value and delay reduction are correlated.

---

16. The error bars denote 95% confidence intervals.
17. In order to evaluate CA-CTA and CTA under the same experimental conditions we ran a new set of experiments using CTA in combination with the driver model detailed in Section 5.2.2.

|  |  | Destination | | | | | | |
|---|---|---|---|---|---|---|---|---|
| **Origin** |  | $O_1$ | $O_2$ | $O_3$ | $O_4$ | $O_5$ | $O_6$ | $O_7$ |
| $O_1$ | CA-CTA | - | 12.22 ± 0.26 | 13.65 ± 0.31 | 25.12 ± 3.40 | 27.13 ± 2.03 | 23.13 ± 0.34 | 13.75 ± 0.11 |
|  | FCFS | - | 11.98 ± 0.31 | 22.89 ± 1.17 | 35.13 ± 1.80 | 43.57 ± 1.89 | 21.35 ± 0.40 | 13.83 ± 0.09 |
|  | %Δ | - | -2.0% | 40.3% | 28.5% | 37.7% | -8.3% | 0.5% |
| $O_2$ | CA-CTA | 12.16 ± 0.21 | - | 10.51 ± 0.14 | 19.58 ± 1.38 | 24.17 ±1.74 | 26.54 ± 0.67 | 22.21 ± 0.37 |
|  | FCFS | 10.15 ± 0.06 | - | 16.50 ± 1.06 | 25.87 ± 1.51 | 31.05 ± 2.03 | 38.09 ± 1.82 | 19.51 ± 0.15 |
|  | %Δ | -19.8% | - | 36.4% | 24.3% | 22.1% | 30.3% | -13.8% |
| $O_3$ | CA-CTA | 15.05 ± 0.69 | 12.51 ± 0.62 | - | 9.01 ± 0.22 | 13.27 ± 0.46 | 18.72 ± 0.68 | 26.76 ± 1.02 |
|  | FCFS | 13.35 ± 0.09 | 9.76 ± 0.03 | - | 12.21 ± 0.62 | 17.64 ± 0.92 | 23.69 ± 6.34 | 31.73 ± 1.36 |
|  | %Δ | -12.7% | -28.2% | - | 26.2% | 24.8% | 21% | 15.7% |
| $O_4$ | CA-CTA | 20.79 ± 1.23 | 18.45 ± 0.93 | 10.52 ± 0.41 | - | 7.32 ± 0.15 | 13.02 ± 1.02 | 23.12 ± 1.53 |
|  | FCFS | 26.94 ± 1.31 | 22.58 ± 1.06 | 13.92 ± 0.82 | - | 10.05 ± 0.48 | 15.74 ± 0.73 | 22.74 ± 0.99 |
|  | %Δ | 22.8% | 18.3% | 24.4% | - | 27.2% | 17.3% | -1.7% |
| $O_5$ | CA-CTA | 24.59 ± 1.10 | 20.82 ± 1.26 | 12.62 ± 0.63 | 7.91 ± 0.48 | - | 10.01 ± 0.28 | 21.88 ± 1.41 |
|  | FCFS | 32.17 ± 1.83 | 30.61 ± 1.70 | 21.54 ± 1.39 | 8.83 ± 0.31 | - | 10.77 ± 0.26 | 17.66 ± 0.52 |
|  | %Δ | 23.6% | 32.0% | 41.4% | 10.4% | - | 7.0% | -23.9% |
| $O_6$ | CA-CTA | 25.08 ± 1.53 | 26.72 ± 0.40 | 18.12 ± 1.26 | 15.78 ± 1.35 | 10.85 ± 0.28 | - | 14.55 ± 0.69 |
|  | FCFS | 22.51 ± 0.40 | 57.01 ± 3.13 | 41.05 ± 2.59 | 24.68 ± 1.62 | 19.02 ± 1.50 | - | 13.73 ± 0.32 |
|  | %Δ | -11.4% | 53.1% | 55.8% | 36.1% | 42.9% | - | -6.0% |
| $O_7$ | CA-CTA | 15.73 ± 0.32 | 24.18 ± 0.52 | 22.12 ± 2.28 | 26.86 ± 2.59 | 16.81 ± 0.99 | 11.43 ± 0.29 | - |
|  | FCFS | 14.31 ± 0.10 | 23.26 ± 0.40 | 56.42 ± 3.01 | 34.99 ± 2.15 | 31.24 ± 2.06 | 12.00 ± 0.20 | - |
|  | %Δ | -9.9% | -3.9% | 60.8% | 23.2% | 46.2% | 4.7% | - |

Table 4: Average travel time in minutes (± 95%CI): CA-CTA vs. FCFS

Table 4 shows the average travel time of the drivers, according to their O-D pairs, when the intersection managers use the CA-CTA mechanism, compared to the FCFS policy. With CA-CTA, there is a net reduction of the average travel time for more than 70% of the O-D pairs if compared to FCFS. Furthermore, at the 30 intersections at which CA-CTA outperforms FCFS, the relative improvement (%$\Delta$) is usually more substantial than the relative losses at the remaining 12 intersections. The travel time reduction is particularly noteworthy for the busy routes $O_6$-$O_2$, $O_6$-$O_3$ and $O_7$-$O_3$ with gains that exceed 50%. On the O-D pairs on which CA-CTA performs worst (especially $O_5$-$O_7$ and $O_3$-$O_2$, with losses of more than 20%) the assignment strategy is not able to sufficiently reduce demand at the intersection, thus considerably increasing the travel time due to the social cost of the combinatorial auction.

## 6. Conclusions

In this article we studied a distributed mechanism for the control and management of a future urban road network, where intelligent autonomous vehicles, controlled by drivers, interact with the infrastructure in order to travel on the links of the network. In this last section we summarise and discuss the main contributions, and we propose some future lines of work.

The first objective was the extension of the reservation-based intersection control system (Dresner & Stone, 2008). We focused on modelling a policy that relied on the theory of combinatorial auctions (Krishna, 2002) to allocate reservations to the drivers. From empirical experimentation, we discovered that the combinatorial auction-based policy guarantees reduced delay to those drivers that value their time the most, i.e., those that submit higher bids. However, this new policy showed that it paid a social cost, in term of greater average delays, especially when traffic demand was high.

The second objective of this work was to go beyond the single intersection setting, and extending the reservation-based model to a network of intersections. Building on the findings reported by Vasirani and Ossowski (2011), we realised that a traffic assignment strategy could make the task of a traffic control policy easier, by better distributing the traffic flow in the network. We studied a market-inspired traffic assignment strategy that tackled the problem from the adaptation perspective. In this model, the intersection managers behaved selfishly, competing with all the others for the supply of the reservations at the intersections. The experimental evaluation showed that in this way the available resources were efficiently allocated to the drivers, generating a more balanced network.

Finally, we combined the competitive strategy for traffic assignment with the auction-based policy for traffic control, in order to develop an adaptive, market-inspired, mechanism for traffic management. The demand-response pricing policy acted on the distribution of vehicles in the network, adapting the reserve price (i.e., the minimum price at which the intersection manager is willing to sell) and generating a system in dynamic equilibrium, where unused intersections became cheaper while highly demanded ones became more expensive. If demand at particularly disputed intersections was lowered by the reserve price fluctuations, the social cost of the auction-based control policy was lowered too (at intersection level). Therefore, a more homogeneous distribution of vehicles over the network led to a better use of network resources, and thus to lower average travel times. In this way, the entire population of drivers was rewarded with lower average travel times and, at the same time, the traffic control policy enforced an inverse relation between bid value and delay, rewarding the drivers that valued the reservations the most with reduced delays.

For future work, other economic models can be implemented, such as continuous double auctions. Furthermore, this work assumed a driver decision making model that exclusively took into consideration the route choice, which was modelled as a utility maximisation problem. In order to capture the inherent complexity of urban traffic systems, it is important to extend and enrich the driver behavioural model. For example, the driver could be implemented as a two layer decision maker, where a reactive, rule-based layer provides short-term decisions about car-following and lane-changing, and a cognitive, BDI-style, layer is in charge of making the more complex decisions such as route choice and departure time selection (Rossetti, Bampi, Liu, Vliet, & Cybis, 2000).

Finally, in this article only interactions between the vehicles and the infrastructure take place. Thus, no collaboration at all is possible between vehicles. Nevertheless, vehicle-to-vehicle communication is receiving great attention from the scientific and engineering community (Biswas, Tatchikou, & Dion, 2006). In particular, vehicle-to-vehicle communication could be used to enrich the action space of a driver, e.g. through the option of dynamically joining or abandoning coalitions of vehicles, based on the idea of platoons (Varaiya, 1993).

## Acknowledgments

This research was partially supported by the Spanish Ministry of Science and Innovation through the project "AT" (CONSOLIDER CSD2007-0022, INGENIO 2010) and "OVAMAH" (TIN2009-13839-C03-02, Plan E).

## References

Adler, J. L., Satapathy, G., Manikonda, V., Bowles, B., & Blue, V. J. (2005). A multi-agent approach to cooperative traffic management and route guidance. *Transportation Research Part B - Methodological*, *39*, 297–318.

Bazzan, A. L. C., & Klügl, F. (Eds.). (2008). *Multi-agent Architectures for Traffic and Transportation Engineering*. IGI-Global.

Biswas, S., Tatchikou, R., & Dion, F. (2006). Vehicle-to-vehicle wireless communication protocols for enhancing highway traffic safety. *IEEE Communications Magazine*, *44*(1), 74–82.

Choy, M. C., Srinivasan, D., & Cheu, R. L. (2003). Cooperative, hybrid agent architecture for real-time traffic control. *IEEE Transactions on Systems, Man, and Cybernetics - Part A*, *33*(5), 597–607.

Clarke, E. H. (1971). Multipart pricing of public goods. *Public Choice*, *11*(1), 17–33.

A Market-Inspired Approach for Intersection Management
Codenotti, B., Pemmaraju, S., & Varadarajan, K. (2004). The computation of market equilibria. *SIGACT News*, *35*(4), 23–37.

da Silva, B. C., Basso, E. W., Bazzan, A. L. C., & Engel, P. M. (2006). Dealing with non-stationary environments using context detection. In *Proceedings of the 23rd International Conference on Machine Learning*, pp. 217–224. ACM.

Dias, M. B., Zlot, R. M., Kalra, N., & Stentz, A. (2006). Market-based multirobot coordination: a survey and analysis. *Proceedings of the IEEE*, *94*(7), 1257–1270.

Dresner, K., & Stone, P. (2008). A multiagent approach to autonomous intersection management. *Journal of Artificial Intelligence Research*, *31*, 591–656.

Gerding, E., McBurney, P., & Yao, X. (2010). Market-based control of computational systems: Introduction to the special issue. *Journal of Autonomous Agents and Multi-Agent Systems*, *21*, 109–114.

Gerlough, D. L., & Huber, M. J. (1975). *Traffic-flow theory*. Transportation Research Board.

Gershenson, C. (2005). Self-organizing traffic lights. *Complex Systems*, *16*, 29–53.

Groves, T. (1973). Incentives in teams. *Econometrica*, *41*(4), 617–631.

Hensher, D. A., & Sullivan, C. (2003). Willingness to pay for road curviness and road type. *Transportation Research Part D - Transport and Environment*, *8*, 139–155.

Hernández, J. Z., Ossowski, S., & García-Serrano, A. (2002). Multiagent architectures for intelligent traffic management systems. *Transportation Research Part C - Emerging Technologies*, *10*(5), 473–506.

Hoos, H. H., & Boutilier, C. (2000). Solving combinatorial auctions using stochastic local search. In *Proceedings of the 17th National Conference on Artificial Intelligence*, pp. 22–29. AAAI Press.

Hunt, P. B., Robertson, D. I., Bretherton, R. D., & Winton, R. I. (1981). Scoot-a traffic responsive method of coordinating signals. Tech. rep., *TRRL Lab. Report 1014*, Transport and Road Research Laboratory, Berkshire.

Ioannou, P., & Chien, C. C. (1993). Autonomous intelligent cruise control. *IEEE Transactions on Vehicular Technology*, *42*(4), 657–672.

Junges, R., & Bazzan, A. L. C. (2008). Evaluating the performance of dcop algorithms in a real world, dynamic problem. In *Proceedings of the 7th International Joint Conference on Autonomous Agents and Multi-Agent Systems*, pp. 599–606. International Foundation for Autonomous Agents and Multiagent Systems.

Krishna, V. (2002). *Auction Theory*. Academic Press.

Krogh, B., & Thorpe, C. (1986). Integrated path planning and dynamic steering control for autonomous vehicles. In *Proceedings of the IEEE International Conference on Robotics and Automation*, pp. 1664–1669.

Lämmer, S., & Helbing, D. (2008). Self-control of traffic lights and vehicle flows in urban road networks. *Journal of Statistical Mechanics: Theory and Experiment*, *2008*(04).

657

Leyton-Brown, K., Shoham, Y., & Tennenholtz, M. (2000). An algorithm for multi-unit combinatorial auctions. In *Proceedings of the 17th National Conference on Artificial Intelligence*, pp. 56–61. AAAI Press.

Meneguzzer, C. (1997). Review of models combining traffic assignment and signal control. *Transportation Engineering*, *123*(2), 148–155.

Nagel, K., & Schreckenberg, M. (1992). A cellular automaton model for freeway traffic. *Journal de Physique I*, *2*(12), 2221–2229.

Ossowski, S., & García-Serrano, A. (1999). Social structure as a computational co-ordination mechanism in societies of autonomous problem-solving agents. In *Intelligent Agents V: Agents Theories, Architectures, and Languages*, Vol. 1555 of *Lecture Notes in Computer Science*, pp. 133–148. Springer.

Papageorgiou, M., Diakaki, C., Dinopoulou, V., Kotsialos, A., & Wang, Y. (2003). Review of road traffic control strategies. In *Proceedings of the IEEE*, Vol. 91, pp. 2043–2067. IEEE.

Robertson, D. I. (1969). Transyt: A traffic network study tool. Tech. rep., *Rep. LR 253*, Road Res. Lab., London.

Rossetti, R. J. F., Bampi, S., Liu, R., Vliet, D. V., & Cybis, H. B. B. (2000). An agent-based framework for the assessment of drivers' decision making. In *Proceedings of the 3rd IEEE Conference on Intelligent Transportation Systems*, pp. 387–392.

Roughgarden, T. (2003). The price of anarchy is independent of the network topology. *Journal of Computer and System Sciences*, *67*(2), 341–364.

Sandholm, T. (2002). Algorithm for optimal winner determination in combinatorial auctions. *Artificial Intelligence*, *135*(1-2), 1–54.

Schepperle, H., & Bohm, K. (2007). Agent-based traffic control using auctions. In *Cooperative Information Agents XI*, Vol. 4676 of *Lecture Notes in Computer Science*, pp. 119–133. Springer.

Schwerdtfeger, T. (1984). Dynemo: A model for the simulation of traffic flow in motorway networks. In *Proceedings of the 9th International Symposium on Transportation and Traffic Theory*, pp. 65–87. VNU Science Press.

Small, K., & Verhoef, E. (2007). *The Economics of Urban Transportation*. Routledge.

Steingrover, M., Schouten, R., Peelen, S., Nijhuis, E., & Bakker, B. (2005). Reinforcement learning of traffic light controllers adapting to traffic congestion. In *Proceedings of the 17th Belgium-Netherlands Conference on Artificial Intelligence*, pp. 216–223.

Treiber, M., Hennecke, A., & Helbing, D. (2000). Congested traffic states in empirical observations and microscopic simulations. *Physical Review E*, *62*(2), 1805–1824.

van Katwijk, R. T., Schutter, B. D., & Hellendoorn, J. (2009). Multi-agent control of traffic networks: Algorithm and case study. In *Proceedings of the 12th International IEEE Conference on Intelligent Transportation Systems*, pp. 316–321.

Varaiya, P. (1993). Smart cars on smart roads: Problems of control. *IEEE Transactions on Automatic Control*, *38*, 195–207.

Vasirani, M., & Ossowski, S. (2009a). Evaluating policies for reservation-based intersection control. In *Proceedings of the 14th Portuguese Conference on Artificial Intelligence*, pp. 39–50.

Vasirani, M., & Ossowski, S. (2009b). A market-inspired approach to reservation-based urban road traffic management. In *Proceedings of the 8th International Joint Conference on Autonomous Agents and Multiagent Systems*, pp. 617–624.

Vasirani, M., & Ossowski, S. (2011). A computational market for distributed control of urban road traffic systems. *IEEE Transactions on Intelligent Transportation Systems*, *12*, 313–321.

Vickrey, W. (1961). Counterspeculation, auctions, and competitive sealed tenders. *Journal of Finance*, *16*, 8–37.

Wiering, M. (2000). Multi-agent reinforcement learning for traffic light control. In *Proceedings of the 17th European Conference on Machine Learning*, pp. 1151–1158.

Wurman, P. R., Wellman, M. P., & Walsh, W. E. (2001). A parametrization of the auction design space. *Games and Economic Behavior*, *35*(1-2), 304–338.

Yen, J. Y. (1971). Finding the k shortest loopless paths in a network. *Management Science*, *17*(11), 712–716.

Zutt, J., van Gemund, A., de Weerdt, M., & Witteveen, C. (2010). Dealing with uncertainty in operational transport planning. In *Intelligent Infrastructures*, pp. 349–375. Springer.